\documentclass[amsmath,amssymb]{revtex4-2}

\usepackage[version=3]{mhchem}
\usepackage[T1]{fontenc} 

\usepackage{soul}
\usepackage{lineno}
\usepackage{fancyhdr}
\usepackage{lastpage}
\usepackage{nomencl}
\usepackage[version=3]{mhchem} 
\usepackage{bm}
\usepackage[usenames,dvipsnames]{color}

\def \be {\begin{equation}}
\def \ee {\end{equation}}
\def \ben {\begin{eqnarray}}
\def \een {\end{eqnarray}}

\usepackage{multirow}
\usepackage{ctable}

\begin{document}

\title{General Chemical Reaction Network Theory for Olfactory Sensing Based on G-Protein-Coupled Receptors : Elucidation of Odorant Mixture Effects and Agonist--Synergist Threshold}

\author{Won Kyu Kim} 
\email{wonkyukim@kias.re.kr}
\affiliation{Korea Institute for Advanced Study, Hoegiro 85, Dongdaemun-gu, Seoul 02455, Korea}

\author{Kiri Choi} 
\affiliation{Korea Institute for Advanced Study, Hoegiro 85, Dongdaemun-gu, Seoul 02455, Korea}

\author{Changbong Hyeon} 
\affiliation{Korea Institute for Advanced Study, Hoegiro 85, Dongdaemun-gu, Seoul 02455, Korea}
\date{\today}

\author{Seogjoo J. Jang}
\email{Seogjoo.Jang@qc.cuny.edu}
\affiliation{Department of Chemistry and Biochemistry, Queens College, City University of New York, 65-30 Kissena Boulevard, Queens, NY 11367 \& PhD programs in Chemistry and Physics, Graduate Center, City University of New York, NY 10016} 
\affiliation{Korea Institute for Advanced Study, Hoegiro 85, Dongdaemun-gu, Seoul 02455, Korea}




\begin{abstract}
This work presents a general chemical reaction network theory for olfactory sensing processes that employ G-protein-coupled receptors as olfactory receptors (ORs). The theory is applicable to general mixtures of odorants and an arbitrary number of ORs. Reactions of ORs with G-proteins, both in the presence and the absence of odorants, are explicitly considered.  A unique feature of the theory is the definition of an odor activity vector consisting of strengths of odorant-induced signals from ORs relative to those due to background G-protein activity in the absence of odorants. It is demonstrated that each component of the odor activity defined this way reduces to a Michaelis--Menten form capable of accounting for cooperation or competition effects between different odorants.  The main features of the theory are illustrated for a two-odorant mixture.  Known and potential mixture effects, such as suppression, shadowing, inhibition, and synergy are quantitatively described. Effects of relative values of rate constants, basal activity, and G-protein concentration are also demonstrated. 
\end{abstract}

\maketitle



\begin{figure}
\includegraphics[width=0.6\textwidth]{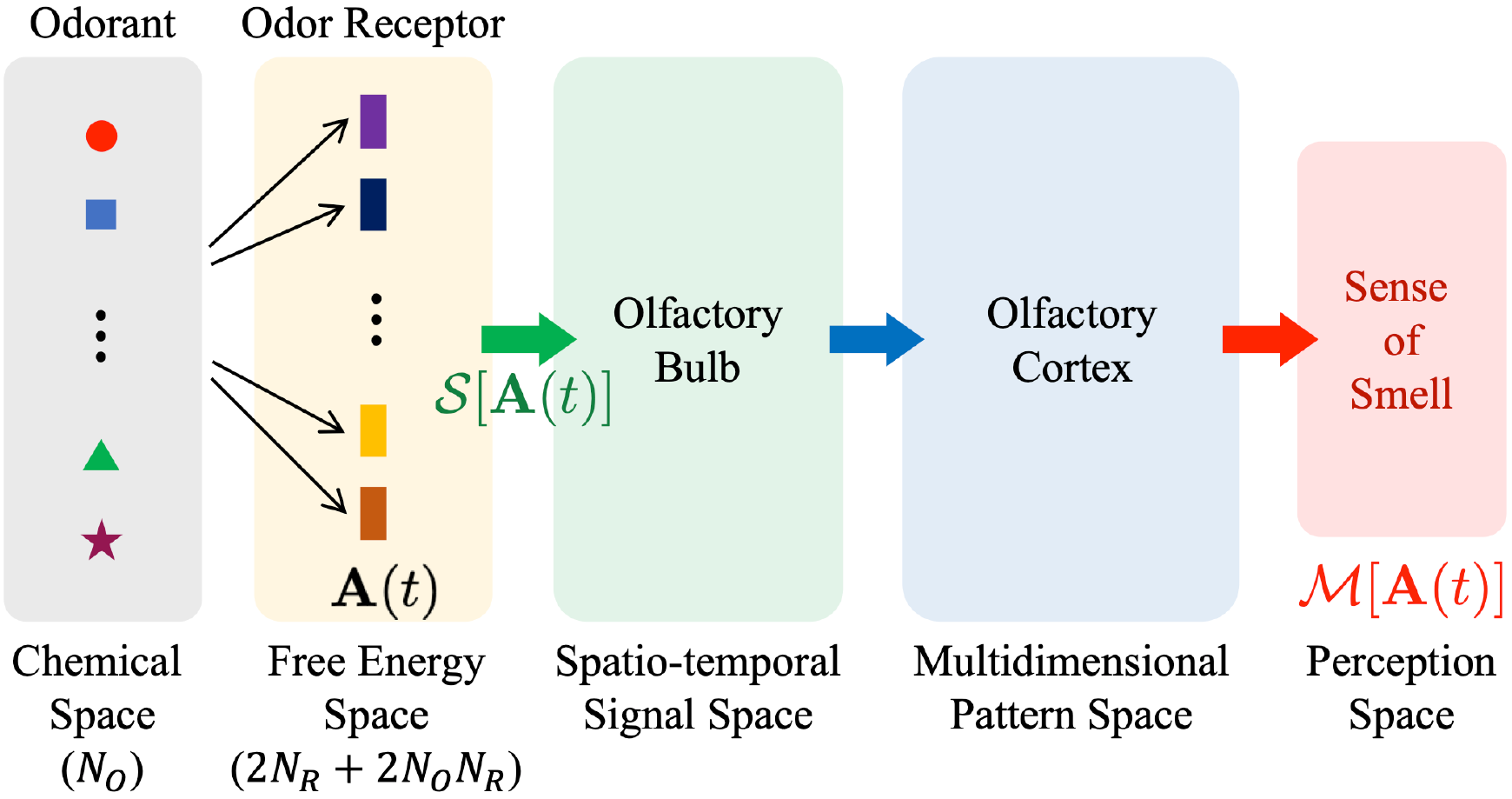}
\caption{Schematic of the olfactory process, where a mixture composed of different odorants is introduced to the olfactory system (see the text for details). Odor Receptor means olfactory receptor (OR).}
\end{figure}
Olfaction, the natural machinery of chemical sensing,~\cite{zazro-brcpc82,demaria-jcb191,Mori-arn34,tromelin-ffj31,block-jafc66,Xu-fee11} in general initiates with the activation of a set of olfactory receptors (ORs) by odorants. Activated ORs then generate electrical signals at the neuronal level~\cite{bhandawat-science308,bhandawat-pnas107,ben-chaim-pnas108,demaria-jcb191,Mori-arn34}, which are further amplified and processed until they are recognized as a sense of smell in the perception space of a brain.  
Details of how ORs become activated and how signals from ORs are processed may vary between organisms, but the majority of the olfaction consists of a few distinctive information processing stages~\cite{tromelin-ffj31} as shown in Fig. 1.
While understanding this level of the mechanism of olfaction is a significant progress that has been achieved through experimental breakthroughs over multiple decades~\cite{zazro-brcpc82,demaria-jcb191,Mori-arn34,tromelin-ffj31,buck-cell65}, many key questions remain unanswered. In particular, there is no well-established and general molecular-level theory capable of explaining the chemical sensitivity, selectivity, and diversity of olfactory sensing\cite{block-jafc66,Xu-fee11,Li-acb9,pshenichnyuk-jpcl9} yet.  Of particular importance is the {\it relationship between odorants-OR reactivity and olfactory codes} that can also be defined for all distinctive mixtures as well.  This letter presents a theory that can serve as such a general framework, upon further refinement and experimental information, by generalizing a recently developed kinetic model of the activation of mammalian ORs~\cite{jang-jpcb121}. 

Our previous kinetic model of olfaction\cite{jang-jpcb121} was developed for mammals that employ the G-protein-coupled receptors (GPCRs) as ORs and is consistent with known information on GPCRs.~\cite{bushdid-jpcl9,demaria-jcb191,demarch-ps24,huang-nature524,sounier-nature524,manglik-cell161,dror-pnas108,lee-bj111,lee-pcb11}   Important features of this model\cite{jang-jpcb121} are as follows. i) The model naturally leads to the Michaelis--Menten type\cite{cornish-bowden-ps4} response to the odorant concentrations, which is fully specified by the reaction coefficients and equilibrium constants of underlying reactions that are in principle measurable. ii) All agonistic, antagonistic, and inverse-agonistic behaviors can be described simply by choosing appropriate rate constants.  This allows for explaining different ORs with one unifying model. iii) The model suggests that quantifying the ${\rm EC_{50}}$ value,\footnote[2]{Defined as the concentration of the odorant where the signal strength becomes half the maximum.} which depends on G-protein concentrations, can offer key information of equilibrium constants.
Overall, the model can describe essential features of dynamics between a single odorant and OR, based on a standard chemical kinetic theory, while suggesting a prescription for quantitative experimental verification.

The generalization of the kinetic model~\cite{jang-jpcb121}  for multiple odorant--OR pairs, as being developed in this letter, is significant considering recent advances in experimental capability to detect responses to mixtures of odorants~\cite{cruz-scirep3,reddy-elife7,singh-pnas116,xu-science368,kurian-ctr383} and relevant theoretical developments~\cite{cruz-scirep3,reddy-elife7,singh-pnas116,kurian-ctr383,marasco-scirep6}. In particular, these research outcomes have established that signals generated by ORs due to mixtures of odorants are far from being additive, and cooperative or inhibitory effects between different odorants are fairly common, instead.  As a result, even if all the information at the individual  odorant-OR level was known, it is not possible to predict the odor codes for the vast majority of mixtures.  It is shown that a general kinetic framework of olfactory sensing for multi-component and multi-OR systems can be constructed so as to elucidate the mechanism and extent of the cooperative effect.

We start our formulation from a standard model of the chemical reaction network but with an arbitrary flux of odorant mixtures. We also explicitly express the reactions between ORs and G-proteins, unlike other existing models. Under the assumption  of the steady-state limit, which is valid given that there is a clear time scale separation between the flux of odorants and reaction processes involving ORs, we obtain a simple expression for the odor activity from the steady-state concentrations of each odorant, equilibrium constants, and the concentrations of the G-protein. 
We show that our theory generalizes the theory utilized by Singh {\it et al.}~\cite{singh-pnas116} and consolidates the effects of competition for G-proteins.

The central assumption of our theory is that  all the chemical information of odorants retrieved by the olfactory system is entirely encoded into their  reactivity with ORs and ensuing changes in the reactivity of ORs with G-proteins.  This assumption is well supported by most primary olfactory sensing mechanisms of vertebrates~\cite{spehr-jn109,bushdid-jpcl9}. Under this assumption, the process leading to the downstream signal processing, which starts from the dissociation of G-proteins from ORs, can be modeled by a set of kinetic equations.   The scope of our theoretical model is to represent these reactions of ORs, together with odorants and G-proteins, and to establish clear relationships between the overall olfactory sensing activity and reaction parameters.

We here introduce terms representing concentrations and probabilities.  We quantify odorants and G-proteins in terms of concentrations, {\it i.e.}, numbers of molecules per unit volume.  On the other hand, we specify ORs of different states in terms of probabilities.  This is because the total number of different ORs, denoted as $N_R$, remains fixed for a given olfactory system. We note that $N_R$ is typically in the range of $100 - 1000$.   For example, $N_R \approx 400$ for humans and $N_R\approx 1,000$ for mice~\cite{niimura-2006,billesbolle2023structural}.

To be more specific, consider an olfactory sensing process where a mixture of different odorants (chemicals) is introduced into the olfactory sensing region (OSR).  Let us assume that the mixture consists of $N_O$ different odorants.  The concentrations and probabilities of odorants are assumed to be time-dependent for now.  Detailed definitions and notations are provided below.
   
\begin{itemize}
\item $C_q(t)$: The concentration of free odorant $q$ within the mixture consisting of $N_O$ odorants,  Thus, $q=1,\cdots, N_O$.  
\item $C_G(t)$: The concentration of free G-proteins.
\item $P_n^u(t)$: Probability of the $n$-th OR in a free state, unbound by any odorant, where $n=1,\cdots, N_R$. 
\item $P_n^G(t)$: Probability of the $n$-th OR in a state forming a complex with a G-protein.  
\item $P_{nq}^{c}(t)$: Probability of the $n$-th OR in a state bound to an odorant $q$ in its active site. The reactivity of this complex with a G-protein is, in general, significantly different from that of the unbound one.
\item $P_{nq}^{cG}(t)$: Probability of the $n$-th OR in a state bound to an odorant $q$ at the active region while also complexed with a G-protein.
\end{itemize}
The sum of all the probabilities for each type of OR has to become unity.  Thus, the following constraint for each of $n=1,\cdots, N_R$ has to be satisfied: 
\be
P_n^u(t)+P_n^G(t)+\sum_{q=1}^{N_O} \left[P_{nq}^c(t)+P_{nq}^{cG}(t) \right] =1.    \label{eq:constraints}
\ee

\begin{figure}
\includegraphics[width=0.5\textwidth]{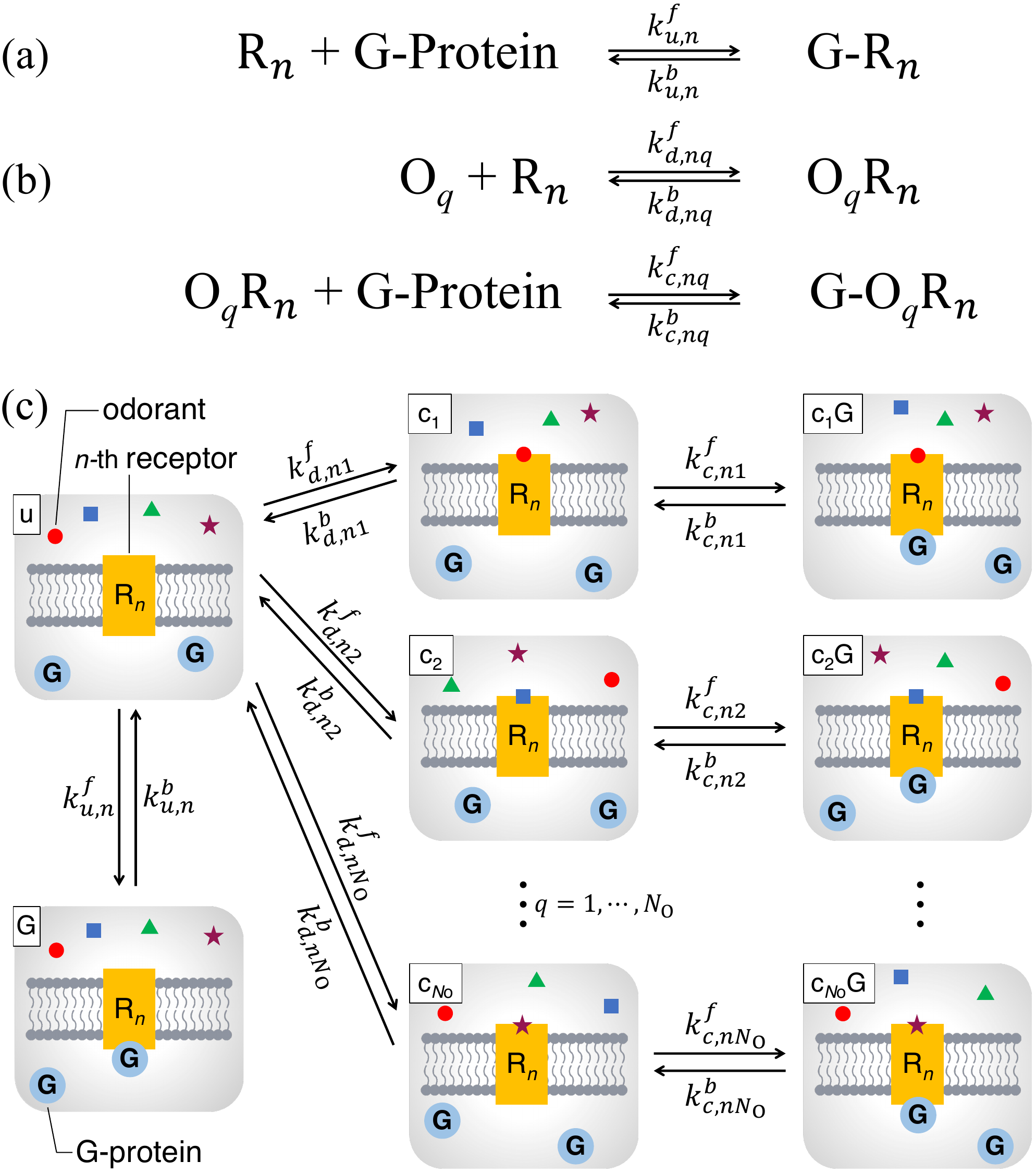}
\caption{\protect\label{reaction-scheme}(a) The reaction of an unbound odor receptor (OR) with a G-protein in the absence of odorants. ${\rm R}_{n}$ represents the $n$-th OR, and  $k_{u,n}^f$ and $k_{u,n}^b$ represent coefficients of forward and backward reactions of unbound ${\rm R}_n$ with a G-protein.   (b) Two reactions of an OR in the presence of odorants.  Coefficients $k_{d,nq}^f$ and $k_{d,nq}^b$ determine the forward and backward reactions of diffusional encounter of an odorant ${\rm O}_q$ and the $n$-th OR, ${\rm R}_n$, which controls the production of the unstable complex ${\rm O_{\textit q} R_{\textit n}}$, followed by relaxation of the odorant into the active site. Coefficients $k_{c,nq}^f$ and $k_{c,nq}^b$ control the forward and backward reaction of the complex formation, namely the odorant-OR-G-protein complex ${\rm G}$-${\rm O_{\textit q} R_{\textit n}}$, respectively. (c) Illustration of complete reaction pathways per receptor, given $N_O$ types of odorants.}
\end{figure}

The overall reaction schemes in the presence and absence of odorants are provided in Fig.~\ref{reaction-scheme}.  
Detailed rate equations, relevant definitions of rate coefficients, and additional terms are provided below.  In this work, we consider idealized chemical reactions and assume that  all reaction rate coefficients are independent of time and concentrations.   

First, the rate equation governing the concentration of each odorant $q=1,\cdots,N_O$ is given by
\ben
&&\frac{\text{d}C_q (t)}{\text{d}t}=I_q(t) - \gamma_q C_q(t)   \nonumber \\
&&\hspace{.2in}+\sum_{n=1}^{N_R} \left[k_{d,nq}^b P_{nq}^{c}(t) - k_{d,nq}^f C_q(t) P_{n}^u (t) \right],  \label{eq:c-q}
\een
where $I_q(t)$ is the time-dependent concentration flux (concentration change per unit time) of the $q$ odorant within a mixture introduced into the OSR.  The coefficient in the second term, $\gamma_q$ is the decay rate of the odorant $q$ due to degradation, deactivation, and escape from the OSR.  In the second line of the above equation, $k_{d,nq}^f$ is the rate coefficient for the forward rate between the $n$-th OR and the odorant $q$ leading to the relaxation of the odorant into the active site of the OR,  and $k_{d,nq}^b$ is the rate for the backward reaction of the same process.   

There are four reactions for each type of OR, involving four different probabilities defined in the last subsection.  
First, the rate equation for the probability of free unbound form of the $n$-th OR is 
\ben
\frac{\text{d}P_{n}^u(t)}{\text{d}t} =\sum_{q=1}^{N_O} \left[ k_{d,nq}^b P_{nq}^{c}(t)  - k_{d,nq}^f C_{q}(t)P_{n}^u(t) \right]  \nonumber \\
-k_{u,n}^f C_G(t)P_{n}^u(t)+k_{u,n}^b P_{n}^G(t),\label{eq:pn-u}
\een
where the terms in the first line are the same as those in the second line in Eq.~(\ref{eq:c-q}), namely rate change due to the dissociation from and the association with each odorant.  As noted above, this rate represents the combined effect of the diffusional encounter/departure between the odorant and the OR and the relaxation of the former into the active site of the latter.    In the second line of the above equation,  $k_{u,n}^f$ is the coefficient of the forward reaction rate of the unbound $n$-th OR forming a complex with a G-protein, and $k_{u,n}^b$ is the coefficient for the backward process, where a G-protein is released from an unbound $n$-th OR.   Note that these two processes are the only reactions that change the probability for the complex between the $n$-th OR and a G-protein.   Thus, for each of $n=1,\cdots, N_R$, 
\be
\frac{\text{d}P_{n}^G(t)}{\text{d}t}  =k_{u,n}^f C_G(t) P_{n}^u(t)-k_{u,n}^b P_{n}^G(t) . \label{eq:pn-g}
\ee

Similarly, the probability for ${\rm O_q R_{n}}$, the $n$-th OR bound with the odorant $q$,  is governed by the following rate equation: 
\ben
\frac{\text{d}P_{nq}^c(t) }{\text{d}t} =k_{d,n q }^f C_{q}(t)P_{n}^u(t) - k_{d,n q }^b P_{nq}^{c}(t)  \nonumber \\
\hspace{.5in}-k_{c,n q }^f C_G(t)P_{nq}^{c}(t)+k_{c, nq}^b P_{nq}^{cG}(t) , \label{eq:pn-q-c}
\een
where $k_{d,nq}^f$ and $k_{d,nq}^b$ are defined through Eqs.~(\ref{eq:c-q}) and (\ref{eq:pn-u}).  In the second line of the above equation, $k_{c,nq}^f$ is the rate coefficient for ${\rm O_q R_n }$ to be complexed with a G-protein, and $k_{c,nq}^b$ is the rate for the backward process.

The probability for G-${\rm O_q R_{n}}$, the $n$-th OR bound with the odorant $q$ and complexed with a G-protein, is governed by the following rate equation:  
\be
\frac{\text{d}P_{nq}^{cG}(t)}{\text{d}t}=k_{c,nq }^f C_G(t)P_{nq}^{c}(t)-k_{c,nq }^b P_{nq}^{cG}(t) , \label{eq:pn-cg}
\ee
where $k_{c,nq }^f$ and $k_{c,nq}^b$ are through Eq.~(\ref{eq:pn-q-c}).

Lastly, the concentration of the G-protein is governed by the following rate equation:
\ben
\frac{\text{d}C_G(t)}{\text{d}t}&=&\sum_{n=1}^{N_R}\Big[-k_{u,n}^f C_G(t) P_n^u(t)+k_{u,n}^b P_n^G(t) \nonumber \\
&&+\sum_{q=1}^{N_O}\left\{-k_{c,nq}^f C_G(t)P_{nq}^c(t)+k_{c,nq}^bP^{cG}_{nq}(t) \right\} \Big]\nonumber \\
&&+I_G (t) -\gamma_G C_G(t), \label{eq:cg-rate}
\een
where all the rate constants have previously been defined through Eqs.~(\ref{eq:pn-u})--(\ref{eq:pn-q-c}), $I_G (t)$ is the flux of G-proteins provided externally, and $\gamma_G$ is the decay rate of the G-protein due to degradation and escape from the OSR. 
We assume that this additional flux and drain mechanisms of G-proteins are much faster than other intermediate rates.

The transduction of the olfactory sensing signal starts from the release of an olfactory signal processing component of the G-protein, known as ${\rm G_\alpha}$ subunit, from G-${\text{O}_{q} \text{R}_{n}}$, the odorant bound OR and G-protein complex.  The released ${\rm G_\alpha}$ subunit then initiates a cascade of events~\cite{rosenbaum-nature459,demaria-jcb191,berbari-cb19,tromelin-ffj31} that involve cyclic adenosine monophosphate (cAMP) messengers, cyclic nucleotide-gated (CNG) ion channels, and ultimately chloride channels.    The depolarization across the membrane of olfactory sensory neurons (OSNs), which primarily involves chloride ions, generates electrical pulses that travel through the OSN and are collected at glomeruli in the olfactory bulb,  which are amplified before they are transferred to projection neurons.  
 
It is known that ORs can still bind to G-proteins in the absence of odorants and generate the same signal-processing components of G-proteins, resulting in non-zero background signals (basal activity).\cite{bhandawat-pnas107,ben-chaim-pnas108}  Thus, we assume that olfactory signal processing can also result from the dissociation of  G-${\rm R}_n$, the complex of an OR and a G-protein without odorant, although at a different rate.  Therefore, we define the following odor activity from the $n$-th OR due to interaction with odorant $q$:
\ben
a_{nq}(t)=\tau_n \left[ k_{c,n q }^b P_{nq}^{cG}(t) +k_{u,n}^b \delta P_{n}^G(t)\right], \label{eq:strength}
\een
where $\tau_n$ is a multiplicative factor characteristic of the $n$-th OR and is  in the unit of time. The physical meaning of $\tau_n$ is that it can be viewed as an ``effective time'' for the duration of signal processing following G-protein release.  Determining this parameter however goes beyond the scope of the chemical network theory presented here.  As will become clear below, it is possible to convert the odor activity into a form that does not explicitly need the value of $\tau_n$. $\delta P_{n}^G(t)$ is the difference between $P_{n}^G(t)$ in the presence of odorants and the absence of odorants.  Then, the odor activity vector for odorant $q$ can be defined as
\be
{\bf a}_q(t)=\left ( \begin{array}{c} a_{1q}(t)  \\ \vdots  \\a_{nq}(t) \\ \vdots \\a_{_{N_R}q}(t) \end{array} \right) . \label{eq:activity}
\ee

While it is possible to define the individual odor activity for each odorant, what is perceived in the actual olfactory sensing is the sum of all contributions of odorants in a given mixture. Then, we can define the following net odor activity vector for a mixture:   
\be
{\bf A} (t)= \left ( \begin{array}{c} A_{1}(t)  \\ \vdots  \\ A_{n}(t) \\ \vdots \\  A_{N_R}(t) \end{array} \right) , \label{eq:net-odor}
\ee
where
\be\label{eq:act}
A_n(t)= \tau_n \left [\sum_{q=1}^{N_O} k_{c,n q }^b P_{nq}^{cG}(t)+ k_{u,n}^b \delta P_{n}^G(t)\right ].
\ee
Thus, the odor perception for a mixture of odorants results from  the mapping of the net odor activity into the perception space, which we denote as ${\mathcal M}({\bf A} (t))$ (see Fig.~1). 

Pending more concrete experimental evidence, it is worth mentioning an important step involved in this mapping process. 
Due to the allosteric nature of the activation of the CNG ion channels, which can combine up to four cAMPs, and other amplification mechanisms, the actual signal strengths that pass through OSNs for a given mixture of odorants are expected to be different from the odor activity defined above.  Since this quantity is experimentally observable, it may be necessary to define the following odor strength for each OR type: 
\be
S_n(t)=H_n(A_n(t)) ,
\ee
where $H_n(x)$ represents an amplification function for each type of OR during neuronal signal generation (see Fig.~1) and is related with the fact that many odor signals can be modeled empirically by Hill function.\cite{bak-2018,reddy-elife7,marasco-scirep6,cruz-scirep3} Possible mechanisms for obtaining such Hill function were provided in our previous work.\cite{bak-2018} In addition, considering that the net outcome of odor signal comes from multiple independent actions of numerous ORs, it is also possible that Poisson-type processes can be employed for the modeling of $H_n(x)$ as well.  Thus, the detailed form of $H_n(x)$ is expected to be dependent on various details of signal transduction and measurements, and  remain an important topic of future theoretical and experimental efforts.

Let us assume that the active olfactory sensing period is preceded by the approach of a steady-state limit, where all the intermediate reactions involving ORs have effectively zero rates on average.  Even in the steady-state limit, the odorant concentration $C_q$ is generally time-dependent because  Eq.~(\ref{eq:c-q}) is not zero due to incoming flux $I_q(t)$ and the decay process with rate $\gamma_q$.  Given that the stationary states of other reaction steps are reached much faster than the time scales of these two processes, the time dependence of $C_q(t_s)$ at the steady-state limit can be determined by
\be
C_q(t_s)\approx \int_0^{t_s} d\tau e^{-\gamma_q (t- \tau)} I_q (\tau)  . \label{eq:C-alpha}
\ee
A similar relation is assumed for $C_G(t_s)$ as well.
Under these assumptions, closed-form expressions for steady-state solutions can be obtained as detailed in Supporting Information (SI). The resulting net odor activity from the interaction between odorant $q$ and the $n$-th OR is 
\ben
&&a_{nq}=\frac{\tau_n C_G}{1+ \sum_{q'=1}^{N_O} C_{q'}f_{nq'} +C_G K_{u,n}}  \left [C_q k_{c,n q }^f  K_{d,nq}  -k_{u,n}^f \frac{\sum_{q'=1}^{N_O} C_{q'}f_{nq'} }{1+ C_G K_{u,n}}\right].  \label{eq:strength-2}
\een
In the above expression, we used the definitions of the association constants and $f_{nq}$ as shown in Table~\ref{tbl:def}.
One important feature of Eq.~(\ref{eq:strength-2}) is that the second term within the square bracket can naturally explain antagonistic and inverse agonistic behaviors.

\begin{table*}[t]
  \caption{Definitions of association constants (ACs) and $f_{nq}$.}
  \label{tbl:def}
  \begin{tabular}{ll}
    \hline
    Symbol & Definition \\
    \hline
    $K_{c,nq}={k_{c,nq }^f}/{k_{c,nq }^b}$ & AC for G-protein binding with odorants  (Eq.~S4 in SI\textsuperscript{\emph{a}})\\
    $K_{d,nq}={k_{d,nq}^f}/{k_{d,nq}^b}$  & AC for odorant binding (Eq.~S5 in SI\textsuperscript{\emph{a}})\\
   $K_{u,n}={k_{u,n}^f}/{k_{u,n}^b}$  &  AC for G-protein binding without odorants (Eq.~S13 in SI\textsuperscript{\emph{a}})\\  
   $f_{nq}=K_{d,nq}\left( 1+ C_G K_{c,nq} \right)$ & A function defined in Eq.~S9 in SI\textsuperscript{\emph{a}}\\  
    \hline
  \end{tabular}

  \textsuperscript{\emph{a}} Supporting Information.
  
\end{table*}

The set of $a_{nq}$s expressed as Eq.~(\ref{eq:strength-2}) completely specifies the odor activity vector  ${\bf a}_q$ defined by Eq.~(\ref{eq:activity}) in the steady-state limit for each pair of odorant and OR. However, what is perceived by the olfactory system is the net odor activity defined by Eq.~(\ref{eq:net-odor}), where each component is given by 
\ben
&&A_n=\frac{\tau_n C_G}{1+ \sum_{q=1}^{N_O} C_{q}f_{nq} +C_G K_{u,n}} \left [\sum_{q=1}^{N_O}C_q \left ( k_{c,n q }^f  K_{d,n q}  -   \frac{ f_{nq} k_{u,n}^f }{1+ C_G K_{u,n}} \right )\right ] . \label{eq:strength-3}
\een   
This is the main result of our theory for the net activity, and can also be rewritten in the Michaelis--Menten (MM) form: 
\ben
A_n&=&\frac{\sum_{q=1}^{N_O} a_{nq}^{0,m} C_q/{\rm EC}_{50}^{nq} }{1+\sum_{q=1}^{N_O} C_q /{\rm EC}_{50}^{nq} }.  \label{eq:strength-4}
\een
In the above expression, $a_{nq}^{0,m}$ defined by Eq.~(S15) in SI is the activity of the $n$-th OR when there is only odorant $q$ in the limit of $C_q \rightarrow \infty$. 
This quantifies the agonistic behavior: $a_{nq}^{0,m}>0$ for agonistic, $a_{nq}^{0,m}=0$ for antagonistic, and $a_{nq}^{0,m}<0$ for inverse-agonistic activities.
Importantly, in principle, the net odorant activity can have negative values when the odorant-triggered activity is weaker than the basal activity defined in Eq.~(\ref{eq:act}).        
Another important quantity ${\rm EC}_{50}^{nq}$ is the corresponding half-maximum activity concentration of odorant $q$ (see Eq.~(S17) in SI). 
Note that in the MM form, the activity is independent of $\tau_n$ because it is rescaled by the maximum activity (see Eq.~(S16) in SI).

Equation (\ref{eq:strength-4}) is similar to the expression suggested by Singh {\it et al.}~\cite{singh-pnas116}
However, we have explicitly derived the net activity $A_n$ in the most general manner.  In addition, our model can account for the effect of basal activity and incorporates G-protein concentrations into $a_{nq}^{0,m}$ and ${\rm EC}_{50}^{nq}$ (see SI for more details).
Since the derived net activity incorporates inverse-agonistic scenarios, it provides quantitative threshold conditions for various OR responses, which we describe in detail below. 

Now, let us consider the case where there are two odorants of types $1$ and $2$, with corresponding steady-state concentrations $C_{1}$ and $C_{2}$.  
Thus, the total concentration of odorants is $C=C_1+C_2$, and the mole fraction of odorant $1$ is $x_1=C_1/C$, thereby yielding the mole fraction of odorant $2$ as $x_2=1-x_1$.

\begin{figure*}
\includegraphics[width=0.99\textwidth]{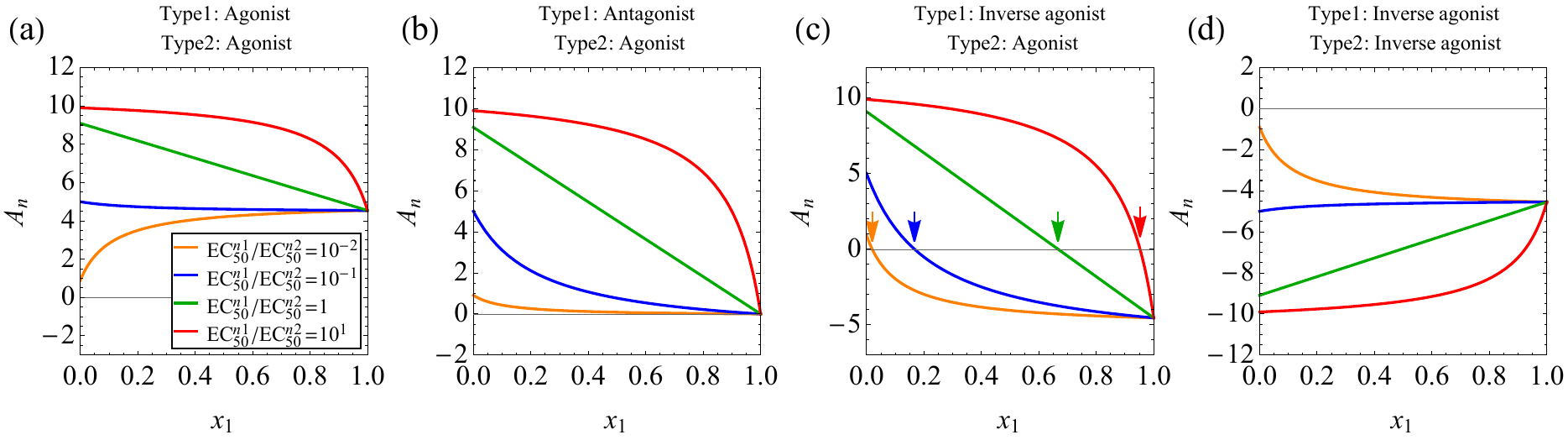}
\caption{\protect\label{fig:MM} Odor activity ${A}_n$ (Eq.~(\ref{eq:an-c-x})) when (a) both types are agonist, (b) type 1: antagonist, type 2: agonist, (c) type 1: inverse agonist, type 2: agonist, and (d) both types are inverse agonist. The arrows in (c) depict the threshold fraction $x_{n1}^{\ast,\text{ago}}$ given in Eq.~(\ref{eq:MMcond}). We use $a_{n1}^{0,m}=5$ and $a_{n2}^{0,m}=10$ for the agonist condition, $a_{n1}^{0,m}=a_{n2}^{0,m}=0$ for the antagonist condition, $a_{n1}^{0,m}=-5$ and $a_{n2}^{0,m}=-10$ for the inverse agonist condition, and $C/{\rm EC}_{50}^{n1}=10$. 
In (a), the red and green curves show ``suppression'', the blue ``overshadowing'', and the orange ``inhibition''. In (b) and (c), all curves show ``suppression''. In (d), the red and green curves show ``suppression'', the blue ``overshadowing'', and the orange ``synergy''.
}
\end{figure*}

According to Eq.~(\ref{eq:strength-4}), the net activity in this case is expressed as 
\be
A_n(C,x_1)=\frac{x_1 \frac{a_{n1}^{0,m}}{{\rm EC}_{50}^{n1}}+ (1-x_1) \frac{a_{n2}^{0,m}}{{\rm EC}_{50}^{n2}} }{\frac{1}{C} + \frac{x_1}{{\rm EC}_{50}^{n1}}+\frac{1-x_1}{{\rm EC}_{50}^{n2}}}.  \label{eq:an-c-x}
\ee 
Depending on the agonist condition for each odorant type (determined by the sign and the magnitude of $a_{n1}^{0,m}$ and $a_{n2}^{0,m}$) as well as ${\rm EC}_{50}^{n1}$ and ${\rm EC}_{50}^{n2}$, the odor activity as a function of $x_1$ changes substantially, and sometimes even qualitatively, as shown in Fig.~\ref{fig:MM}.
More comprehensive and detailed calculation results are provided in sections IV and V of SI.

Figure~\ref{fig:MM}(a) illustrates the odor activity when both odorant types are agonistic but type 2 is two-fold stronger ($a_{n1}^{0,m}=5$ and $a_{n2}^{0,m}=10$). 
Depending on the ratio ${\rm EC}_{50}^{n1}/{\rm EC}_{50}^{n2}$, $A_n$ decreases or increases, converging to a value close to $a_{n1}^{0,m}=5$ as $x_1$ increases to unity.
This is because we chose the parameter value $C/{\rm EC}_{50}^{n1}=10$, meaning that the odorant concentration is sufficiently large to reach $a_{n1}^{0,m}$ for type 1.
Note that the parameter ratio, 
\be
\frac{{\rm EC}_{50}^{n1}}{{{\rm EC}_{50}^{n2}}}=\frac{K_{d,n2}(1  + C_G K_{c,n2} )}{K_{d,n1}(1  + C_G K_{c,n1} )},\label{eq:ECratio}
\ee
is dictated by the odorant binding ratio via $K_{d,n2}/K_{d,n1}$ for small $C_G$, and approaches $(K_{d,n2} K_{c,n2})/(K_{d,n1} K_{c,n1})$ for large $C_G$.
As ${\rm EC}_{50}^{n1}/{\rm EC}_{50}^{n2}$ increases, the activity in the limit of vanishing $x_1$ approaches $a_{n2}^{0,m}=10$, which fulfills the saturation concentration for both types (see the red curve for ${{\rm EC}_{50}^{n1}}/{{{\rm EC}_{50}^{n2}}}=10$).
We observe rich phenomena of activity in Fig.~\ref{fig:MM}(a) depending on different values of ${{\rm EC}_{50}^{n1}}/{{{\rm EC}_{50}^{n2}}}$: The red and green curves exhibit “suppression,” the blue “overshadowing,” and the orange “inhibition.”
Note that the choice of values for $a_{n1}^{0,m}=5$ and $a_{n2}^{0,m}=10$ above, although somewhat arbitrary, are good examples representing the case of two comparable but different odorants, for which rich cooperative and competitive effects are expected.  For more comprehensive model calculations including other values of parameters, please refer sections IV and V of SI (see Figs. S2-S4 in SI).

While keeping the identical agonist condition for odorant type 2 ($a_{n2}^{0,m}=10$), we now consider the antagonist condition ($a_{n1}^{0,m}=0$) for type 1, as shown in Fig.~\ref{fig:MM}(b). 
As $x_1$ increases, the activity indeed decreases to zero, where the decay characteristic is determined by ${\rm EC}_{50}^{n1}/{\rm EC}_{50}^{n2}$ and exhibits a “suppression” of activities.

Interestingly, as shown in Fig.~\ref{fig:MM}(c), when the odorant type 1 is inverse agonistic ($a_{n1}^{0,m}=-5$) and the type 2 is agonistic ($a_{n2}^{0,m}=10$), we find a crossover (sign change) of the activity $A_n$ as $x_1$ changes.
This indicates that the agonist condition of the mixture can be tuned (from agonistic to inverse agonistic and vice versa) by changing a fraction of one odorant type. 
The crossover occurs at the threshold fraction (obtained by solving $A_n=0$ in Eq.~(\ref{eq:an-c-x})
 \be\label{eq:MMcond}
 x_{n1}^{\ast,\text{ago}} = \frac{1}{1-\frac{a_{n1}^{0,m} {\rm EC}_{50}^{n2}}{a_{n2}^{0,m} {\rm EC}_{50}^{n1}}},
 \ee
which satisfies $0 < x_{n1}^{\ast} \leq 1$ if and only if ${a_{n1}^{0,m}}/{a_{n2}^{0,m}} \leq 0$, meaning that the odor activity crossover occurs only in a mixture of agonistic and inverse agonistic odorant types. 
Although the activity features “suppression” responses for all ${\rm EC}_{50}^{n1}/{\rm EC}_{50}^{n2}$, the agonistic condition is largely dependent on $x_1$ [see the arrows in Fig.~\ref{fig:MM}(c)].  We expect this result will be particularly useful for further quantitative studies of odorant-dependent OR response and widespread inhibition. 
Recently, Inagaki {\it et al.} have found that the OR response can be largely modulated by changing a fraction of one odorant type in a mixture of agonists and inverse agonists~\cite{Inagaki-2020}. The OR response can undergo excitatory-to-inhibitory transitions by changing a fraction of inverse agonists in a mixture (see Fig.~4B in Ref.~\cite{Inagaki-2020}). In addition, Pfister {\it et al.} have found the inhibitory-to-excitatory transitions of OR response by increasing an agonist fraction in a mixture of agonists and inverse agonists (see Fig.~7B in Ref.~\cite{Pfister-2020}). These findings are consistent with $A_n$ in Fig.~\ref{fig:MM}(c) at ${\rm EC}_{50}^{n1}/{\rm EC}_{50}^{n2}=10$ (red curve), the threshold of which can be quantitatively determined by Eq.~(\ref{eq:MMcond}).   
Furthermore, the result enables us to define the following vector for the threshold fraction of odorant type $q$:
\be
{\bf x}_q^{\ast,\text{ago}}=\left ( \begin{array}{c} x_{1q}^{\ast,\text{ago}}  \\ \vdots  \\x_{nq}^{\ast,\text{ago}} \\ \vdots \\x_{N_R q}^{\ast,\text{ago}} \end{array} \right), \label{eq:thresx}
\ee
which provides the transition threshold profile of odor receptors for given odor types in a mixture. 

Figure~\ref{fig:MM}(d) shows $A_n$ when both types are inverse agonistic ($a_{n1}^{0,m}=-5$ and $a_{n2}^{0,m}=-10$), which is equal to $-A_n$ of the agonist case shown in Fig.~\ref{fig:MM}(a). We observe that the red and green curves exhibit “suppression”, the blue “overshadowing”, and the orange “synergy”.

Comparing the above net odor activity with those for single-odorant cases can classify different behaviors of a mixture, specifically, the synergistic behavior. 
Therefore, we define a relative activity
\be
\delta A_n(C,x_1)=A_n(C,x_1)-a_{n1}^0(Cx_1)-a_{n2}^0 [C(1-x_1)] , \nonumber
\ee
where $a_{n1}^0(Cx_1)$ is the odor activity for odorant $1$ at concentration $Cx_1$ in the absence of odorant $2$, and $a_{n2}^0 [C(1-x_1)]$ is the odor activity for odorant $2$ at concentration $C(1-x_1)$ in the absence of odorant $1$. Then, we find 
\ben
\delta A_n(C,x_1) = \frac{-C^2\frac{x_1(1-x_1)}{{\rm EC}_{50}^{n1} {\rm EC}_{50}^{n2}}}{1+C \left(\frac{x_1}{{\rm EC}_{50}^{n1}}+\frac{1-x_1}{{\rm EC}_{50}^{n2}} \right)} \left ( \frac{a_{n1}^{0,m}}{1+C \frac{x_1}{{\rm EC}_{50}^{n1}}}  +\frac{a_{n2}^{0,m}}{1+C \frac{1-x_1}{{\rm EC}_{50}^{n2}}} \right ).  \label{eq:dA}
\een 
\begin{figure*}[t]
\includegraphics[width=0.99\textwidth]{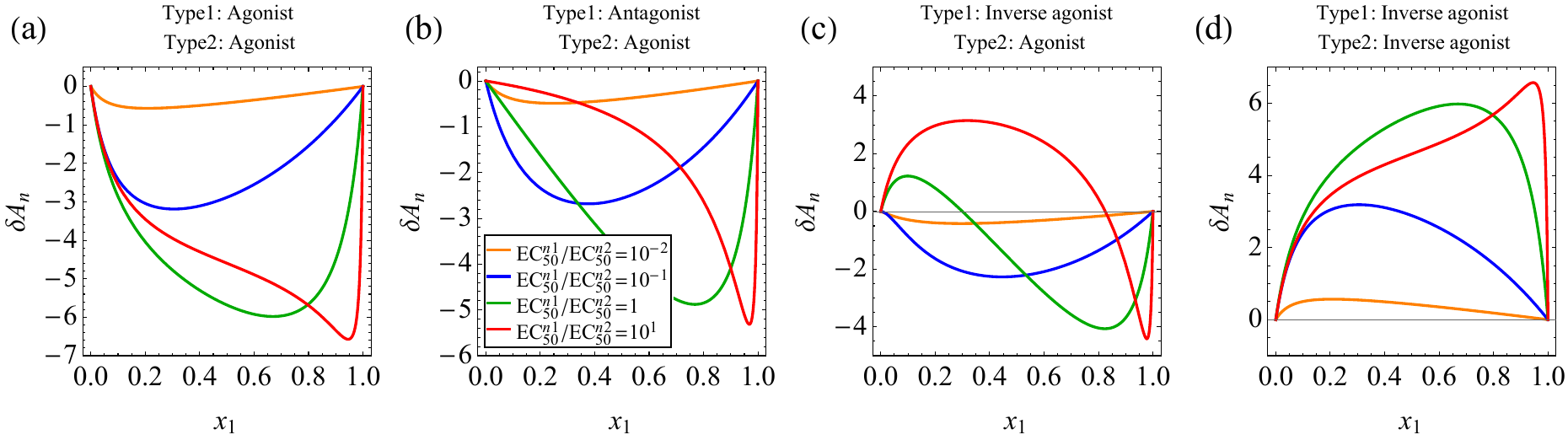}
\caption{\protect\label{fig:dA} Relative odor activity ${\delta A}_n$ given by Eq.~(\ref{eq:dA}) when (a) both types are agonist, (b) type 1: antagonist, type 2: agonist, (c) type 1: inverse agonist, type 2: agonist, and (d) both types are inverse agonist. We use $a_{n1}^{0,m}=5$ and $a_{n2}^{0,m}=10$ for the agonist condition, $a_{n1}^{0,m}=a_{n2}^{0,m}=0$ for the antagonist condition, $a_{n1}^{0,m}=-5$ and $a_{n2}^{0,m}=-10$ for the inverse agonist condition, and $C/{\rm EC}_{50}^{n1}=10$. }
\end{figure*}

Figure~\ref{fig:dA}(a) depicts $\delta A_n$ when both odorant types are agonistic ($a_{n1}^{0,m}=5$ and $a_{n2}^{0,m}=10$), which is the same condition as given in Fig.~\ref{fig:MM}(a).
We find that $\delta A_n$ is always negative. Thus, the net odor activity is always smaller than the direct algebraic sum of independent and individual activities. This suggests that a synergistic enhancement is impossible (anti-synergistic).

When the odorant type 1 is antagonistic and type 2 is agonistic ($a_{n1}^{0,m}=0$ and $a_{n2}^{0,m}=10$), we find a similar anti-synergistic condition [see Fig.~\ref{fig:dA}(b)].  However, when odorant type 1 is inverse agonistic ($a_{n1}^{0,m}=-5$) and type 2 is agonistic ($a_{n2}^{0,m}=10$), we find a large variation of $\delta A_n$ [Fig.~\ref{fig:dA}(c)].
Depending on $x_1$ and ${\rm EC}_{50}^{n1}/{\rm EC}_{50}^{n2}$, the net activity can be synergistic ($\delta A_n>0$) or anti-synergistic ($\delta A_n<0$). The threshold synergistic condition is found as
\be
x_{n1}^{\ast, \text{syn}}=\frac{1+\frac{a_{n2}^{0,m}{\rm EC}_{50}^{n2}}{a_{n1}^{0,m} C}+\frac{{\rm EC}_{50}^{n2}}{C}}{1-\frac{a_{n2}^{0,m} {\rm EC}_{50}^{n2}}{a_{n1}^{0,m} {\rm EC}_{50}^{n1}}}, \label{eq:x1syn}
\ee 
which satisfies $0 < x_{n1}^{\ast, \text{syn}} \leq 1$ if and only if ${a_{n1}^{0,m}}/{a_{n2}^{0,m}} \leq 0$.
The odor activation is synergistic in the range of $0 < x_1 < x_{n1}^{\ast, \text{syn}}$, where the range becomes larger as ${\rm EC}_{50}^{n1}/{\rm EC}_{50}^{n2}$ increases and vice versa for the anti-synergistic range.
The change of $\delta A_n(x_1)$ is substantial, particularly resulting in qualitative difference for large ${\rm EC}_{50}^{n1}/{\rm EC}_{50}^{n2}$, where the slope of the decaying curve at the crossover point becomes sharper.  This implies that a small change in $x_1$ can significantly modulate the synergistic condition.
  
\begin{figure}[t]
\includegraphics[width=0.45\textwidth]{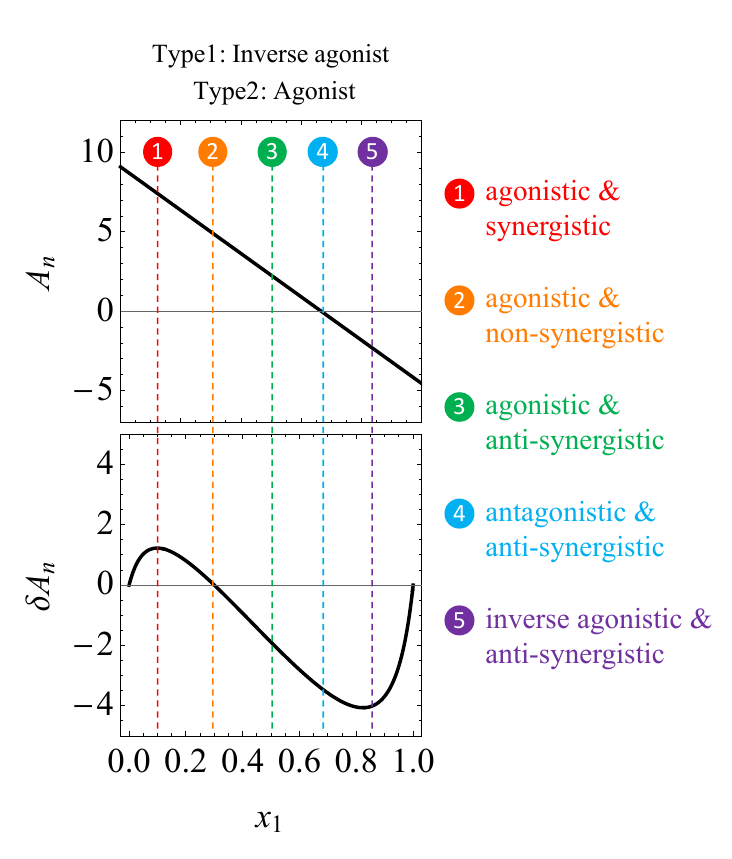}
\caption{\protect\label{fig:synerg} Agonist--synergist condition for a mixture of inverse agonists (type 1: $a_{n1}^{0,m}=-5$) and agonists (type 2: $a_{n2}^{0,m}=10$) at ${\rm EC}_{50}^{n1}/{\rm EC}_{50}^{n2}=1$ and $C/{\rm EC}_{50}^{n1}=10$, which depends on $x_1$. Five different scenarios are depicted by numeric indices. 1: Agonistic and synergistic, 2: Agonistic and non-synergistic, 3: Agonistic and anti-synergistic, 4: Antagonistic and anti-synergistic, and 5: Inverse agonistic and anti-synergistic.}
\end{figure}

In Fig.~\ref{fig:dA}(d), when both types are inverse agonistic ($a_{n1}^{0,m}=-5$ and $a_{n2}^{0,m}=-10$), $\delta A_n(x_1)$ is identical to $-\delta A_n$ in the agonist case shown in Fig.~\ref{fig:dA}(a).  Note that in our model the synergistic activity occurs only when the mixture contains an inverse agonist. 
This means that the synergistic effect in a GPCR-based signal downstream is dictated by the number of inhibitory encounters (odorants, ligands, or drugs).
Moreover, the synergistic threshold vector,
\be
{\bf x}_q^{\ast, \text{syn}}=\left ( \begin{array}{c} x_{1q}^{\ast, \text{syn}}  \\ \vdots  \\x_{nq}^{\ast, \text{syn}} \\ \vdots \\x_{N_R q}^{\ast, \text{syn}} \end{array} \right), \label{eq:thresxsyn}
\ee
provides a synergistic threshold profile of odor receptors given the odor types in a mixture. 
This result offers insights into a quantitative understanding of OR-dependent response and synergy criteria.  
Recently, Inagaki {\it et al.} have found that depending on ORs, the odor activity response by an identical mixture can be considerably different---``suppression'' or ``synergy'' (see Fig.~6B in Ref.~\cite{Inagaki-2020}).
What is notable is that the synergistic activity has been found from the OR response by a mixture of agonists and inverse agonists in the experiment, which is consistent with our theoretical model ({\it e.g.}, see scenario 1 in Fig.~\ref{fig:synerg}).

It is worthwhile to explain several important points relevant to our main results.
For mammalian olfactory GPCR signaling, the concentration of G-proteins in an OSN is known to be $\sim 1~\text{mM}$~\cite{li-pnas119-2022,rodieck-98}.
In addition, a success probability of odorant-triggered signal downstream from GPCR to G-protein is found to be very low ($\sim 10^{-4}$)~\cite{li-pnas119-2022}.
This suggests that the ratio ${\rm EC}_{50}^{n1}/{\rm EC}_{50}^{n2}$ given in Eq.~(\ref{eq:ECratio}) for olfactory sensing becomes
\be
\frac{{\rm EC}_{50}^{n1}}{{\rm EC}_{50}^{n2}}=\frac{K_{d,n2}}{K_{d,n1}},
\ee
where the key factor determining the agonist--synergist condition is the odorant-binding (association) constant.

Figure~\ref{fig:synerg} shows the agonist--synergist condition depending on the mole fraction $x_1$ of odorant type 1 based on the net activity $A_n$ and the relative activity $\delta A_n$ for a mixture of inverse agonists (type 1: $a_{n1}^{0,m}=-5$) and agonists (type 2: $a_{n2}^{0,m}=10$) at ${\rm EC}_{50}^{n1}/{\rm EC}_{50}^{n2}=1$ and $C/{\rm EC}_{50}^{n1}=10$.
Various scenarios are possible: 1: Agonistic and synergistic, 2: Agonistic and non-synergistic, 3: Agonistic and anti-synergistic, 4: Antagonistic and anti-synergistic, and 5: Inverse agonistic and anti-synergistic conditions.
In general, these conditions are determined by the threshold space $({\bf x}_q^{\ast, \text{ago}},{\bf x}_q^{\ast, \text{syn}})$ given by Eqs.~(\ref{eq:MMcond}) and (\ref{eq:thresxsyn}), which can be measured in experiments in terms of the Michaelis--Menten parameters.

Interestingly, agonist--synergist modulation becomes significant when there are inhibitory ligands, i.e., inverse-agonistic odorants, in the mixture. A noteworthy crossover between the net and the relative activities occurs only when a pair of agonists and inverse agonists compose the two-odorant mixture.  
We found the agonist--synergist condition can be tuned by changing a mole fraction of one odorant in a mixture.  
This is in line with previous observations of ``competitive antagonism'' and ``widespread modulation'',~\cite{reddy-2018, xu_2023} where antagonists or inverse agonists~\cite{deMarch_2020} in the odorant mixture modulate the signal downstream. 

Various theoretical models~\cite{reddy-2018, xu_2023, rospars-2008,marasco-scirep6,cruz-scirep3,bak-2018}, mostly based on the Hill equation as an empirical formula, have been used to understand the odorant-dependent modulation of receptor reaction by mixtures.
Recently, Singh {\it et al.} presented the ``competitive-binding'' model~\cite{singh-pnas116} based on the Michaelis--Menten equation similar to what we derived previously~\cite{jang-jpcb121} and is extended here, and showed possible widespread modulation of ORs in an equimolar mixture ($x_1$=$x_2$=1/2).
Our model provides further information on microscopic parameters, such as G-protein concentrations, association constants, and more importantly, the basal activity, and can serve as a general framework for understanding the model by Singh {\it et al.}\cite{singh-pnas116} 

It should be noted that experimental observations point towards a broad range of basal activities across different types of OSNs~\cite{Reisert-2010,Connelly-2013,Inagaki-2020}.
We find that $C_G {K}_{u,n}$ is crucial for determining the basal activity.
For low basal activity, the probability for G-proteins to bind to an OR in the absence of odorants (Eq.~(S22) in SI) is small, which vanishes as $C_G {K}_{u,n} \rightarrow 0$.  Instead, for large $C_G {K}_{u,n}$, the basal activity is high.
Therefore, ${\rm EC}_{50}^{nq}$ (Eq.~(S17) in SI) turns out to depend on the basal activity, while the ratio ${\rm EC}_{50}^{n1}/{\rm EC}_{50}^{n2}$ for a mixture is independent of the basal activity. This finding may be useful for quantitative analyses of olfactory sensing with mixtures.

In conclusion, in this work, we generalized the rate equations for multi-odorant and multi-OR reactions by explicitly incorporating the effects of odorant binding, G-protein binding, and basal activity.
The resultant analytical solutions yield the net activity and relative activity expressed in terms of the Michaelis--Menten equation, with corresponding microscopic biochemical contributions as functions of $x_1$ and $C_G$.  Our theory provides mechanisms and quantitative criteria for the important effects of odorant mixtures. In addition, our main theoretical results can be used to construct feature space of odorants based on their reactivity with ORs.   Thus, our theory, along with recent experimental progress to objectively measure odor codes at the neuronal level,\cite{Xu-fee11} can be utilized for the development of more efficient and experimentally verifiable Machine Learning approaches for olfaction since the space of chemical rates and equilibrium constants can serve as an effective feature space for modeling olfactory codes.

While our theory describes the effects of competitive multi-odorant binding and G-protein binding for olfactory GPCR signaling, the underlying model can also be adapted to provide a general framework for elucidating similar GPCR-based ligand-receptor-effector reaction processes with ligand mixtures, e.g., gustatory sensing~\cite{McLaughlin-1992} and behavioral regulation with a mixture of neurotransmitters~\cite{McCorvy-2015}.    
It is important to note that our assumption that combines a diffusional encounter of odorant and OR, as well as the relaxation of the former into an active site of the latter as a single-step process, might be  drastic simplification. Nonetheless, considering that this process is likely to be faster than the diffusional encounter and the reaction of the unbound OR with a G-protein, it is reasonable to represent the two processes by a single rate equation. Even in the case where these assumptions do not hold, it is still possible to explicitly represent the two processes into two separate rate equations, which introduces an additional equilibrium constant in the steady-state limit.
We expect that further refinement of our model with an increasing number of experimental data is possible. The present results are based on the steady-state solutions, which are important but yet describe only a fraction of our kinetic model.
For longer time scales, one can incorporate the dynamics of concentrations of odorants and G-proteins, as given by Eqs.~(\ref{eq:c-q}) and (\ref{eq:cg-rate}), which will provide a more comprehensive perspective to understand the kinetics of olfactory sensing throughout a wide temporal range.

\section*{acknowledgement}
SJJ acknowledges major support from the National Science Foundation (CHE-1900170) and Korea Institute for Advanced Study (KIAS) for partial support through KIAS scholar program.  Additional support came from KIAS Individual Grants CG076002 (WKK), CG077002 (KC), and CG035003 (CH). 
We thank the Center for Advanced Computation in KIAS for providing the computing resources.

%

\setcounter{equation}{0}
\setcounter{figure}{0}
\setcounter{table}{0}
\setcounter{page}{1}

\renewcommand{\theequation}{S\arabic{equation}}
\renewcommand{\thefigure}{S\arabic{figure}}
\renewcommand{\thetable}{S\arabic{table}}
\renewcommand\thepage{S\arabic{page}}

\section{Supporting Infomation}

\subsection{Solution of rate equations in the steady-state limit} 
We consider the steady-state limit where all the rate equations involving the intermediates in the main text become zero.  Although the concentrations can still be time-dependent, for notational convenience, we drop all the concentrations with such time dependence.

Let us first start from the conditions that Eqs.~5 and 6 become zero. 
From the condition that Eq.~6 is zero, we obtain
\be
k_{c,nq}^f P_{nq}^c C_G =k_{c,nq}^b P_{nq}^{cG} . \label{eq:pncg-st}
\ee
Combining the above condition with the condition that Eq.~5 is zero and rearranging terms, we find that
\be
 P_{nq}^{c}=\frac{k_{d,nq}^f}{k_{d,nq}^b} C_q P_n^u  . \label{eq:pnacs-1}
 \ee
Employing the above expression in Eq.~\ref{eq:pncg-st}, we find that 
\be
P_{nq}^{cG}=\frac{k_{c,nq}^f}{k_{c,nq}^b}\frac{k_{d,nq}^f}{k_{d,nq}^b} C_q C_GP_n^u .  \label{eq:pnacg-1}
\ee 

Now let us define the following equilibrium constants 
\ben
&&K_{c,nq}=\frac{k_{c,nq }^f}{k_{c,nq }^b}  , \label{eq:Kc}\\
&&K_{d,nq}=\frac{k_{d,nq}^f}{k_{d,nq}^b} . \label{eq:Kd}
\een
With the above definitions, Eqs.~\ref{eq:pnacs-1} and \ref{eq:pnacg-1} can be expressed as 
\ben
& P_{nq}^c&= K_{d,nq} C_q P_n^u,\label{eq:pnac-2} \\
& P_{nq}^{cG}&= K_{d,nq} K_{c,nq}C_q C_G P_n^u .  \label{eq:pnacg-2}
\een

Combining Eqs.~\ref{eq:pnac-2} and \ref{eq:pnacg-2}, we obtain
\be
P_{nq}^c+P_{nq}^{cG} =C_q f_{nq}  P_n^u , \label{eq:sum-p-q}
\ee
where
\be
f_{nq}=K_{d,nq}\left[1+ C_G K_{c,nq} \right ]  . \label{eq:f-q} 
\ee
Note that $f_{nq}$ defined by Eq.~\ref{eq:f-q} is a linear function of the concentration of G-protein, $C_G$, which is determined by the specific physiological condition in the olfactory regions.  
All other parameters in $f_{nq}$ are fully determined by equilibrium coefficients.

Employing Eq.~\ref{eq:sum-p-q} in the normalization condition, Eq.~1, we obtain
\be
P_n^u =\frac{1-P_n^G}{1+\sum_{q=1}^{N_O} C_q f_{nq}} .  \label{eq:pnu-pnug}
\ee 
The above equation provides a relationship between $P_n^u$ and $P_n^G$,  probabilities of completely unbound OR and G-protein-only bound OR in the steady-state limit.  Combining this information with the condition that Eq.~4 is also zero in the steady-state limit, we obtain the following expression
\ben
P_n^G=\frac{C_G K_{u,n}}{ 1+\sum_{q=1}^{N_O} C_q f_{nq} + C_G K_{u,n} } ,  \label{eq:png-2} \\
P_n^u= \frac{1}{1+ \sum_{q=1}^{N_O}  C_q f_{nq} +C_G K_{u,n}} , \label{eq:pnu-2}
\een
where   
\be
K_{u,n}=\frac{k_{u,n}^f}{k_{u,n}^b} . \label{eq:kun}
\ee

\subsection{Single odorant case}
We here consider the case where $N_O=1$ with only odorant $q$, for which Eq.~\ref{eq:strength-2} reduces to
\ben
a_{nq}^0&=&\frac{\tau_n C_GC_q}{1+ C_{q}f_{nq} +C_G K_{u,n}}\nonumber \\
&&\times \left [ k_{c,n q }^f  K_{d,nq}  -    \frac{ f_{n,q} k_{u,n}^f }{1+ C_G K_{u,n}} \right ]  . \label{eq:strength-ampha}
\een    
In the limit where $C_q \rightarrow \infty$, the above quantity reaches the following maximum:
\ben
a_{nq}^{0,m}&=&\frac{\tau_n C_G}{f_{nq}}\left [ k_{c,n q }^f  K_{d,nq}  -   \frac{ f_{nq} k_{u,n}^f }{1+ C_G K_{u,n}} \right ]  .\label{eq:strength-alpha-m}
\een   
Therefore, 
\ben
\frac{a_{nq}^0}{a_{nq}^{0,m}}&=&\frac{C_q f_{nq}}{1+ C_{q}f_{nq} +C_G K_{u,n}} \nonumber \\
&\equiv&\frac{C_q}{{\rm EC}_{50}^{nq}+C_q} , 
\een
where ${\rm EC}_{50}^{nq}$ is the concentration of odorant $q$ for which its odor activity with the $n$-th OR becomes half of its maximum and is expressed as
\ben 
{\rm EC}_{50}^{nq}&=&\frac{1+C_GK_{u,n}}{f_{nq}} \nonumber \\
&=&\frac{1+C_G K_{u,n}}{K_{d,nq}(1  + C_G K_{c,nq} )} . \label{eq:ec-50-alpha}
\een

Employing Eq.~\ref{eq:ec-50-alpha}, we can express $f_{nq}$ in terms of ${\rm EC}_{50}^{nq}$ as follows:
\be
f_{nq}=\frac{1+C_GK_{u,n}}{{\rm EC}_{50}^{nq}} . \label{eq:fn-alpha} 
\ee
Using this expression in Eq.~\ref{eq:strength-alpha-m} and rearranging terms, we find the following expression:
\be
\tau_nC_G k_{c,nq}^f K_{d,nq} =\frac{a_{nq}^{0,m} (1+C_G K_{u,n})}{{\rm EC}_{50}^{nq}-K_{nq}} , \label{eq:tau-n}
\ee
where 
\be
K_{nq}=\frac{k_{u,n}^f}{k_{c,nq}^f K_{d,nq} }. \label{eq:kn-def}
\ee

\begin{figure*}
\includegraphics[width=0.98\textwidth]{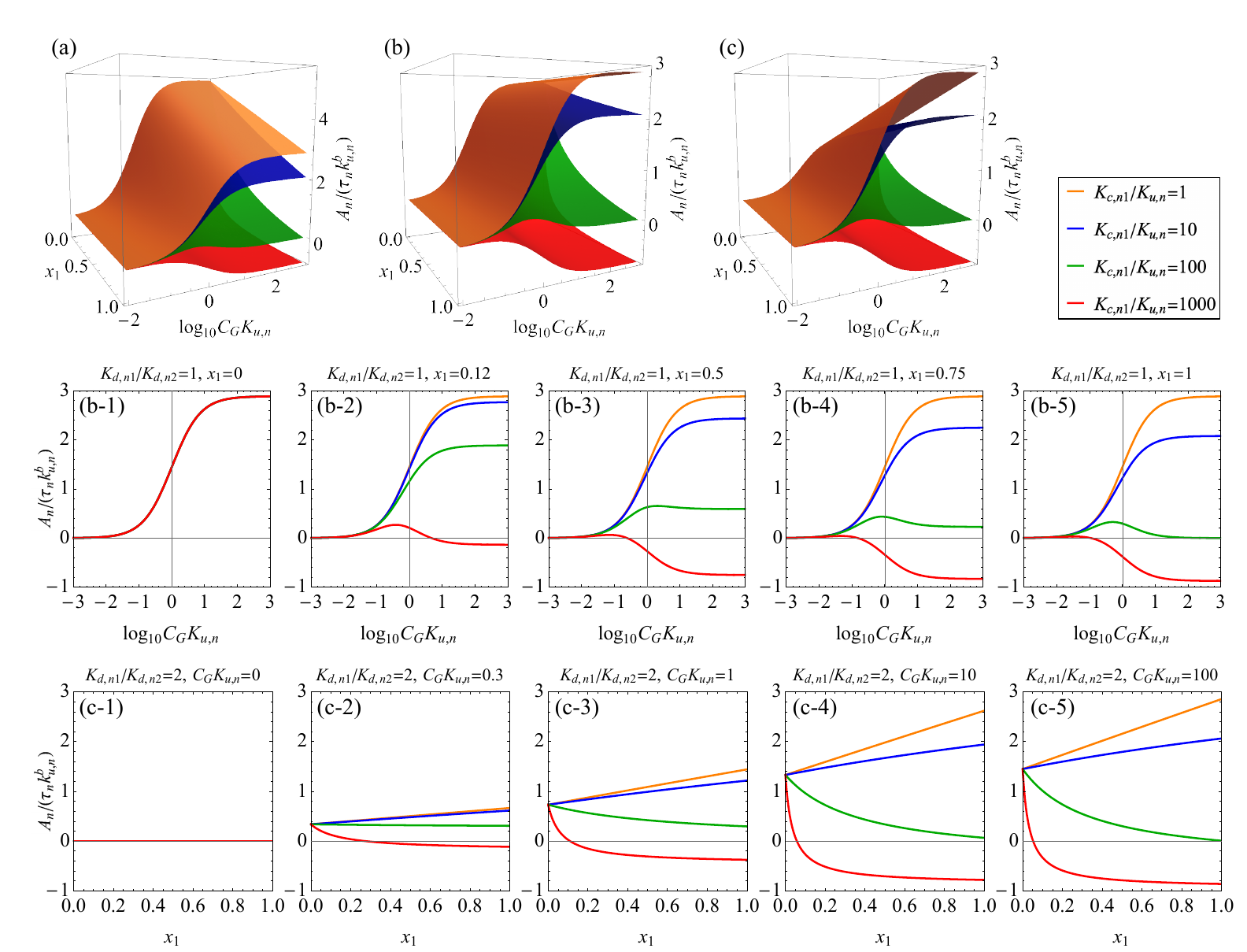}
\caption{\label{An3d}Odor activity $\tilde{A}_n (\tilde{C}_G, x_1)$ in Eq.~\ref{eq:An_dimless} at (a) ${K}_{d,n1}/{K}_{d,n2}=0.5$, (b) ${K}_{d,n1}/{K}_{d,n2}=1$, and (c) ${K}_{d,n1}/{K}_{d,n2}=2$ for different values of ${K}_{c,n1}/{K}_{u,n}$. Middle row: Odor activity $\tilde{A}_n (\tilde{C}_G)$ at ${K}_{d,n1}/{K}_{d,n2}=1$ [panel (b)] for (b-1) $x_1=0$, (b-2) $x_1=0.12$, (b-3) $x_1=0.5$, (b-4) $x_1=0.75$, and (b-5) $x_1=1$. 
Bottom row: Odor activity $\tilde{A}_n (x_1)$ at ${K}_{d,n1}/{K}_{d,n2}=2$ [panel (c)] for (c-1) $\tilde{C}_G=0$, (c-2) $\tilde{C}_G=0.1$, (c-3) $\tilde{C}_G=1$, (c-4) $\tilde{C}_G=10$, and (c-5) $\tilde{C}_G=100$. 
We use $k^f_{c,n1}/k^f_{u,n}=k^f_{c,n2}/k^f_{u,n}=100$, ${K}_{c,n2}/{K}_{u,n}=1$, and $CK_{d,n1}=0.03$.}
\end{figure*}

\subsection{Derivation of $a_n$ and $A_n$}

Under the steady-state condition with Eq.~13, simple expressions for steady-state solutions can be obtained as detailed in Supporting Information.   The resulting solutions for probabilities, dropping the explicit dependences on $t_s$ for notational convenience, are given by Eqs.~S11 and S12. Equation~S12 can be combined with Eq.~S7 to determine the steady-state value of $P_{nq}^{cG}$ that is needed for the calculation of the first term in the odor activity defined by Eq.~8.  Thus, we obtain the following expression:
\be
P_{nq}^{cG}=\frac{K_{d,nq}K_{c,nq} C_q C_G}{1+ \sum_{q=1}^{N_O}  C_q f_{nq} +C_G K_{u,n}} ,  \label{eq:pnacg-3}
\ee 
where $K_{c,nq}$, $K_{d,nq}$, $f_{nq}$, and $K_{u,n}$ are defined by Eqs.~S4, S5, S9, and S13, respectively.

For the calculation of the second term in Eq.~8, we also need to determine the steady-state value of $P_n^G$ in the absence of odorants.  This can be obtained by setting  $C_q=0$ for all $q$ in Eq.~S11, which results in the following expression:
\be
P_{n,0}^G=\frac{C_G K_{u,n}}{1+ C_G K_{u,n}} . \label{eq:PGn0}
\ee
Subtracting this from Eq.~S11, we obtain
\be
\delta P_{n}^G=-\frac{ (\sum_{q=1}^{N_O} C_q f_{nq})C_G K_{u,n}}{ (1+\sum_{q=1}^{N_O} C_q f_{nq}+ C_G K_{u,n} ) (1+ C_G K_{u,n} )} . \label{eq:delta-png-0}
\ee

Using Eqs.~\ref{eq:pnacg-3} and \ref{eq:delta-png-0} in Eq.~8 result in the following steady-state expression for the odor activity from the interaction between odorant $q$ and the $n$-th OR:
\ben
&&a_{nq}=\frac{\tau_n C_G}{1+ \sum_{q'=1}^{N_O} C_{q'}f_{nq'} +C_G K_{u,n} } \nonumber \\
&& \times \left [C_q k_{c,n q }^f  K_{d,nq}  -k_{u,n}^f \frac{\sum_{q'=1}^{N_O} C_{q'}f_{nq'} }{1+ C_G K_{u,n}}\right ]  .\label{eq:strength-2}
\een

\subsection{Net odor activity depending on $C_G$ and $x_1$}

For a two-odorant mixture, the net odor activity $A_n$ in Eq.~15 is rewritten in a dimensionless form:
\be\label{eq:An_dimless}
\tilde{A}_n=\frac{ x_1 \tilde{K}_{d,n1}  \left( \tilde{k}^f_{c,n1} - \Gamma_1 \right) + (1-x_1) \tilde{K}_{d,n2}  \left( \tilde{k}^f_{c,n2} - \Gamma_2 \right) }{(1+{1}/{\tilde{C}_G}) [1+x_1 \tilde{K}_{d,n1} \Gamma_1 + ( 1- x_1 ) \tilde{K}_{d,n2} \Gamma_2]},
\ee
where we use $\Gamma_\alpha=(1+\tilde{C}_G \tilde{K}_{c,n\alpha})/(1+\tilde{C}_G)$, $\tilde{A}_n={A_n}/({\tau_n k^b_{u,n}})$, $\tilde{C}_G=C_G K_{u,n}$, $\tilde{K}_{d,nq}=K_{d,nq} C$, $\tilde{k}^f_{c,nq}={k}^f_{c,nq}/k^f_{u,n}$, $\tilde{K}_{c,nq}={K}_{c,nq}/K_{u,n}$ with $q=1,2$. For $x_1=1$, $\tilde{A}_n$ reduces to the activity for the single odorant case (Eq.~20 in Ref.~\cite{jang-jpcb121}). 

In Fig.~\ref{An3d}, we show $\tilde{A}_n$ as a function of $\tilde{C}_G$ and $x_1$, which features a rich landscape depending on the odorant binding constant ratio (${K}_{d,n1}/{K}_{d,n2}$) and relative G-protein binding ratio for odorants of type 1 (${K}_{c,n1}/{K}_{u,n}$). Here, we use $\tilde{k}^f_{c,n1}=\tilde{k}^f_{c,n2}=100$, $\tilde{K}_{c,n2}=1$, and $\tilde{K}_{d,n1}=0.03$, following the parameters considered previously~\cite{jang-jpcb121}.
This means that we keep the agonist condition for odorants of type 2 ($\tilde{k}^f_{c,n2}>\tilde{K}_{c,n2}$) and we vary $\tilde{K}_{c,n1}$, which changes the agonist condition for odorants of type 1.
These conditions are depicted by color codes in Fig.~\ref{An3d}, in which starting from orange to red, odorants of type 1 become agonists (orange, blue), antagonists (green), and inverse agonists (red), while odorants of type 2 are always agonistic.

When the odorant binding ratio is ${K}_{d,n1}/{K}_{d,n2}=0.5$, as shown in Fig.~\ref{An3d}(a), the activity is overall a decreasing function of $x_1$. This is because odorants of type 2 bind to OR two times faster than type 1, thus the activity is suppressed by increasing $x_1$, which becomes less and inverse agonistic as $\tilde{K}_{c,n1}/{K}_{u,n} $ increases. 
In Fig.~\ref{An3d}(b), when the odorant binding ratio is unity (${K}_{d,n1}/{K}_{d,n2}=1$), the $x_1$-dependency of activity is determined by the agonist condition of odorants of type 2: For $\tilde{K}_{c,n1}=1$ (orange surface) that is actually the same as $\tilde{K}_{c,n2}=1$, two odorant types are indistinguishable, thus the activity is independent of $x_1$ and features the s-curve function of G-protein concentration [see orange curves in panels (b-1)--(b-5)].
For $\tilde{K}_{c,n1}>1$, however, the activity becomes a decreasing function of $x_1$ because of the fraction of agonistic type 2 decreases. 
When odorants of type 2 bind two times faster (${K}_{d,n1}/{K}_{d,n2}=10$), as shown in Fig.~\ref{An3d}(c), the activity shows intriguing $x_1$-dependent behavior [see panels (c-1)--(c-5)] , which can be an increasing function of $x_1$ under the agonist condition for odorant type 1, otherwise it is a decreasing function of $x_1$.

The various behavior of odorant activity arises from the competition between two odorant types with different agonist conditions.  
Up to $\tilde{K}_{c,n1}=100$, we find that the activity is positive for all $x_1$ and $C_G$. 
For $\tilde{K}_{c,n1}/{K}_{u,n} =1000$ (red), the activity is positive only in a certain limited range of $x_1$ and $C_G$ [see the crossover of the red lines in Figs.~\ref{An3d}(b-1)-(d-5)]. This can be further generalized as the odor activity of a binary mixture is limited by the threshold fraction $x_{n1}^\ast$:
\be\label{eq:x_1ast}
x_{n1}^\ast (C_G) = \frac{1}{1 - \frac{ \tilde{K}_{d,n1}  \left( \tilde{k}^f_{c,n1} - \Gamma_1 \right) }{ \tilde{K}_{d,n2}  \left( \tilde{k}^f_{c,n2} - \Gamma_2 \right) }},
\ee
above which the olfactory signal subtracted by the basal activity, Eq.~12, becomes negative (inverse agonistic), provided that type $2$ is agonistic (vice versa, $x_{n1}^\ast$ is the threshold to be agonistic when type 2 is inverse agonistic).

This agonistic competition in the mixture is well reflected in Figs.~\ref{An3d}(b-1)--(c-5), which are cross-sections of $A_n (x_1,\tilde{C}_G)$ in Fig.~\ref{An3d}(b) cut for various fixed values of $x_1$.
The typical s-curve feature of $A_n (C_G)$ for single-type agonistic odorants [Fig.~\ref{An3d}(b-1)] changes dramatically when mixed with other odorants of different types with various agonist conditions [Figs.~\ref{An3d}(b-2)--(b-4)]. Depending on $\tilde{C}_G$ and the mixing fraction $x_1$, the activity can be monotonically agonistic, nonmonotonically agonistic, and even nonmonotonically inverse agonistic.
Interestingly, when a small fraction of inverse agonistic odorants is mixed [the red curve in panel (b-2)]. As $\tilde{C}_G$ increases from zero, the activity is maximized and becomes negative as $\tilde{C}_G$ increases further, showing the role of G-protein concentration dramatically controlling the olfactory sensing. 
This behavior is not found for the single odorant case [panels (b-1) and (b-5)], which is the unique feature arising from a mixture.
Another interesting result is found when mixing antagonistic odorants with agonistic odorants [the green curves in panels (b-3) and (b-4)], where the activity is also maximized at an optimal $\tilde{C}_G$ and the mixture becomes agonistic as $\tilde{C}_G$ increases further.  

In addition, to explicitly show the effects of the threshold condition $x_{n1}^\ast$ in Eq.~\ref{eq:x_1ast} on the activity, we show $A_n (x_1,\tilde{C}_G)$ in Fig.~\ref{An3d}(c) cut for various fixed values of $\tilde{C}_G$ in Figs.~\ref{An3d}(c-1)--(c-5).
As $\tilde{C}_G$ increases, the crossover of the activity is clearly seen in the red curves for the inverse-agonistic odorants of type 1, of which the crossover point is quantified by Eq.~\ref{eq:x_1ast}.
 
The obtained theoretical result may provide insight into a quantitative analysis of the competitive effect of an odorant mixture.   

\subsection{Connection to basal activity}

\begin{figure*}
\includegraphics[width=0.99\textwidth]{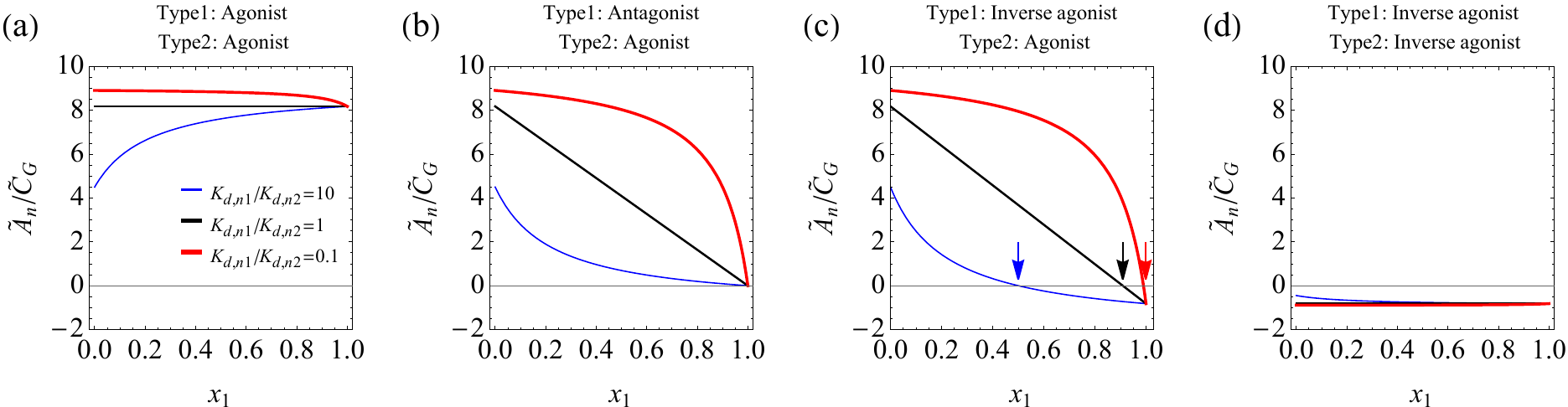}
\caption{\label{fig:lowbasalactivity} Odor activity $\tilde{A}_n / \tilde{C}_G$ given in Eq.~\ref{eq:Anlowbasalact}, under the low basal activity condition ($\tilde{C}_G \ll 1$) for (a) both types agonist, (b) type1: antagonist, type2: agonist, (c) type1: inverse agonist, type2: agonist, and (d) both types inverse agonist. The arrows in (c) depict the threshold fraction $x_{n1}^{\ast, \text{low}}$ given in Eq.~\ref{eq:lowthres}. For an odorant of type $\alpha$, we use $\tilde{k}^f_{c,n\alpha}=10$ for the agonist condition, $\tilde{k}^f_{c,n\alpha}=1$ for the antagonist condition, $\tilde{k}^f_{c,n\alpha}=0.1$ for the inverse agonist condition, and $\tilde{K}_{d,n1}=10$. }
\end{figure*}

\begin{figure*}
\includegraphics[width=0.99\textwidth]{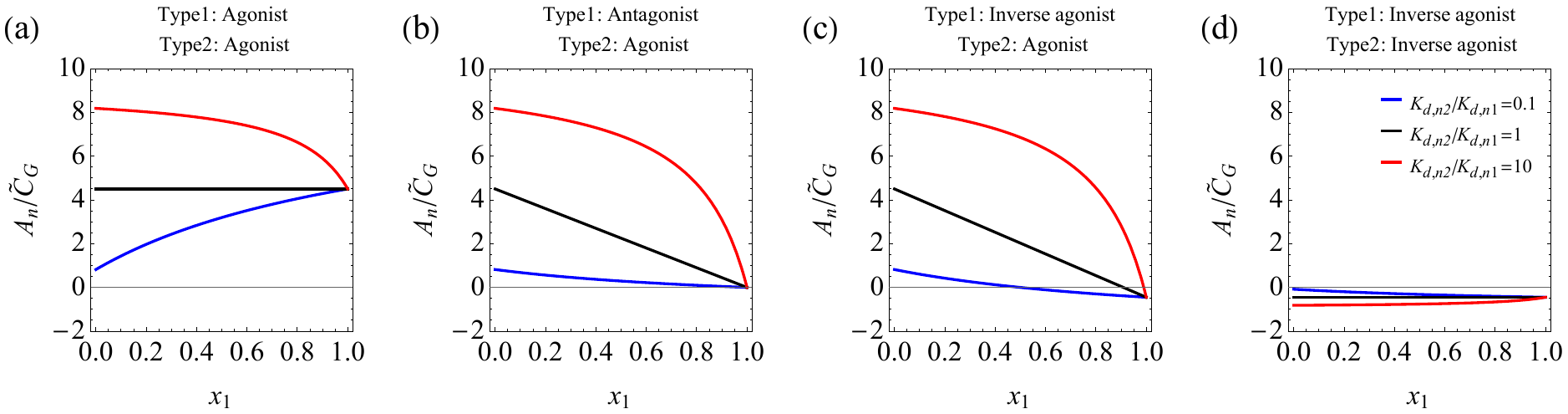}
\caption{\label{fig:lowbasalactivity2} Odor activity $\tilde{A}_n / \tilde{C}_G$ under low basal activity ($\tilde{C}_G \ll 1$), for (a) both types agonist, (b) type1: antagonist, type2: agonist, (c) type1: inverse agonist, type2: agonist, and (d) both types inverse agonist. For an odorant of type $\alpha$, we use $\tilde{k}^f_{c,n\alpha}=10$ for the agonist condition, $\tilde{k}^f_{c,n\alpha}=1$ for the antagonist condition, $\tilde{k}^f_{c,n\alpha}=0.1$ for the inverse agonist condition, and $\tilde{K}_{d,n1}=1$. }
\end{figure*}

\begin{figure*}
\includegraphics[width=0.99\textwidth]{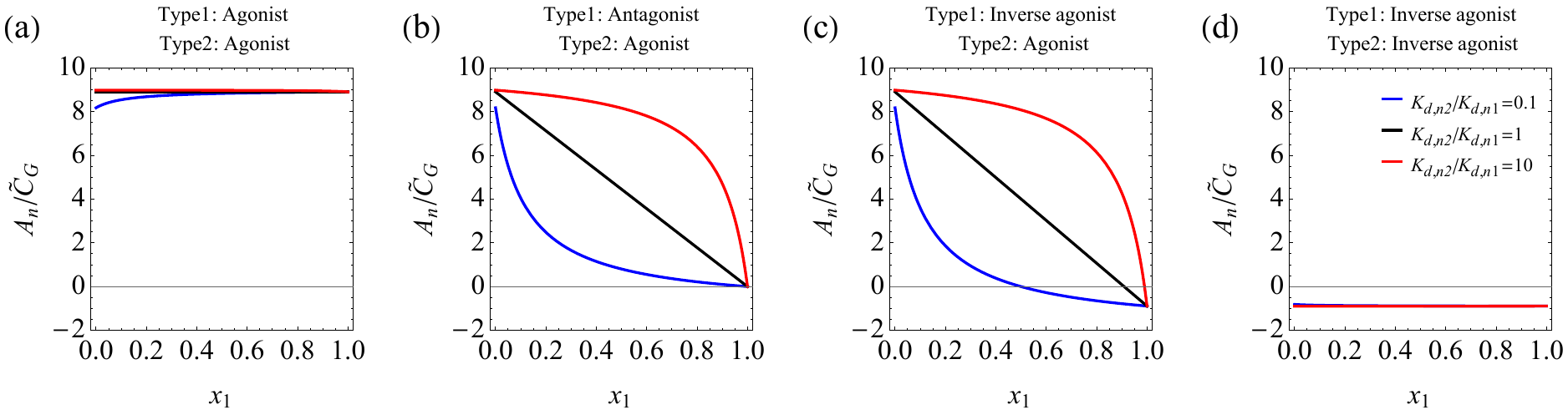}
\caption{\label{fig:lowbasalactivity3} Odor activity $\tilde{A}_n / \tilde{C}_G$ under low basal activity ($\tilde{C}_G \ll 1$) for (a) both types agonist, (b) type1: antagonist, type2: agonist, (c) type1: inverse agonist, type2: agonist, and (d) both types inverse agonist. For an odorant of type $\alpha$, we use $\tilde{k}^f_{c,n\alpha}=10$ for the agonist condition, $\tilde{k}^f_{c,n\alpha}=1$ for the antagonist condition, $\tilde{k}^f_{c,n\alpha}=0.1$ for the inverse agonist condition, and $\tilde{K}_{d,n1}=100$. }
\end{figure*}

We now consider two limiting cases of the activity, focusing on a connection to the basal activity.

First, we focus on the low basal activity condition, under which the binding (or association) constant $(K_{u,n})$ for G-proteins (of constant concentration) to bind to an odorant-unbound OR (u $\rightarrow$ G process in Fig.~2) is small, i.e., $\tilde{C}_G \ll 1$.
We then find a leading order expression of the activity,
\be\label{eq:Anlowbasalact}
{ \tilde{A}_n=\tilde{C}_G \frac{ x_1 \tilde{K}_{d,n1}  \left( \tilde{k}^f_{c,n1} - 1 \right) + (1-x_1) \tilde{K}_{d,n2}  \left( \tilde{k}^f_{c,n2} - 1 \right) }{1 + x_1 \tilde{K}_{d,n1} + ( 1- x_1 ) \tilde{K}_{d,n2} }.}
\ee
Note that a key quantity that determines the activity is $\tilde{k}_{c,n\alpha}^f$, which is the binding rate of G-proteins to bind to an odorant-bound OR relative to $k_{u,n}^f$.  
One thus finds that the agonist condition for single-type odorants is $\tilde{k}^f_{c,n\alpha} > 1$.
However, for a binary mixture of odorants, we find a criterion for the mixture to be agonist,
\be\label{eq:lowBAcondition}
\frac{ {K}_{d,n1} (\tilde{k}^f_{c,n1} - 1) }{ {K}_{d,n2} (\tilde{k}^f_{c,n2}-1) } > \frac{x_1 - 1}{x_1}.
\ee 
The right-hand side of the above equation is always negative or zero, therefore the inequality holds always when the sign of $\tilde{k}^f_{c,n1} - 1$ and $\tilde{k}^f_{c,n2} - 1$ is the same., i.e., when both types are agonistic or inversely agonistic.
This means that for all $x_1$, $A_n > 0$ when both types are agonists and $A_n < 0$ when both types are inverse agonists.
Interestingly, a crossover of $A_n (x_1)$ can occur when one type is an agonist and the other is an inverse agonist, which controls the activity by $x_1$ with the threshold condition,
\be\label{eq:lowthres}
x_{n1}^{\ast, \text{low}} = \frac{1}{1-\frac{ {K}_{d,n1} (\tilde{k}^f_{c,n1} - 1) }{ {K}_{d,n2} (\tilde{k}^f_{c,n2}-1) }},
\ee
determining of the sign of activity.

Figure~\ref{fig:lowbasalactivity} shows the activity $\tilde{A}_n(x_1)/\tilde{C}_G$ for different odorant binding rate ratios ${K}_{d,n2}/{K}_{d,n1}$ and different agonist conditions.
When both odorant types are agonistic, as shown in Fig.~\ref{fig:lowbasalactivity}(a), the mixture is also agonistic and the sign of $A_n$ is always positive for all $x_1$. 
Depending on the odorant binding ratio ${K}_{d,n2}/{K}_{d,n1}$, the activity can be a decreasing or increasing function of $x_1$, which eventually converges to the value at ${K}_{d,n2}/{K}_{d,n1}=1$ as $x_1$ increases up to unity (the single component limit).
This tendency of activity is also found in Fig.~\ref{fig:lowbasalactivity}(b) when odorants of type 1 are antagonistic and type 2 are agonistic, which decreases to zero as $x_1$ increases.
Note that for ${K}_{d,n2}/{K}_{d,n1}=1$, the activity becomes a linear function of $x_1$.
When odorants of type 1 are inverse agonistic and type 2 are agonistic, as shown in Fig.~\ref{fig:lowbasalactivity}(c), we find the crossover of activity as $x_1$ increases. The threshold fraction $x_{n1}^{\ast, \text{low}}$ given in Eq.~\ref{eq:lowthres} is depicted by the arrows in Fig.~\ref{fig:lowbasalactivity}(c).
When both odorant types are inverse agonistic, as shown in Fig.~\ref{fig:lowbasalactivity}(d), the mixture is also inverse agonistic and the sign of $A_n$ is always negative for all $x_1$. 
We also show $\tilde{A}_n(x_1)/\tilde{C}_G$ in Figs.~\ref{fig:lowbasalactivity2} and \ref{fig:lowbasalactivity3} for $\tilde{K}_{d,n1}=1$ and $\tilde{K}_{d,n1}=100$ respectively, of which the overall tendency is similar to Fig.~\ref{fig:lowbasalactivity}.

Figures~\ref{fig:lowbasalactivity2} and \ref{fig:lowbasalactivity3} show the activity with the same parameters except for $\tilde{K}_{d,n1}=1$ and $\tilde{K}_{d,n1}=100$, respectively, of which the overall tendency is similar.

\begin{figure}
\includegraphics[width=0.3\textwidth]{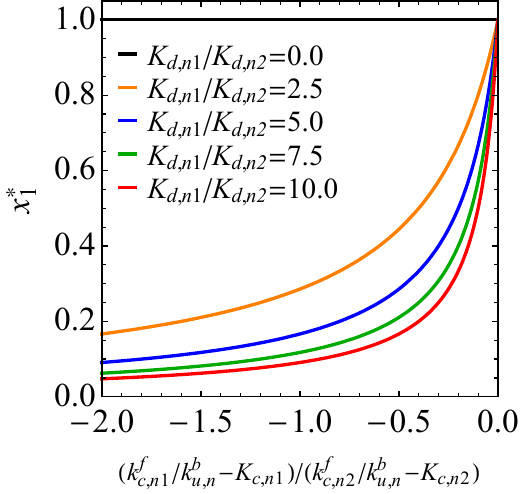}
\caption{\label{fig:x1}Threshold odorant fraction $x_{n1}^{\ast, \text{high}}$ in Eq.~\ref{eq:x_1ast_highbasal} as a function of ${ ({k}^f_{c,n1}/{k}^b_{u,n}-{K}_{c,n1} ) }/{ ({k}^f_{c,n2}/{k}^b_{u,n}-{K}_{c,n2} ) }$ for different $ {K}_{d,n1}/{K}_{d,n2}$.}
\end{figure}

Now suppose that the G-protein-to-OR binding constant $K_{u,n}$ is still small but the concentration of G-protein is exceedingly high, yielding $\tilde{C}_G \gg 1$.
This condition can also apply to a case of high basal activity condition when $K_{u,n} \gg1$, under which the signal downstream can be strong even without odorant-binding events. 
In the limit of $\tilde{C}_G \rightarrow \infty$, we find
\be
{\scriptstyle \tilde{A}_n=\frac{ x_1 \tilde{K}_{d,n1}  \left( \tilde{k}^f_{c,n1} - \tilde{K}_{c,n1} \right) + (1-x_1) \tilde{K}_{d,n2}  \left( \tilde{k}^f_{c,n2} - \tilde{K}_{c,n2} \right) }{1 + x_1 \tilde{K}_{d,n1} \tilde{K}_{c,n1} + ( 1- x_1 ) \tilde{K}_{d,n2} \tilde{K}_{c,n2} },}
\ee
which yields the agonist criterion,
\be\label{eq:cri_largeCg}
\frac{ {K}_{d,n1} ( \tilde{k}^f_{c,n1} - \tilde{K}_{c,n1} ) }{{K}_{d,n2} (  \tilde{k}^f_{c,n2} - \tilde{K}_{c,n2} ) } > \frac{x_1-1}{x_1}.
\ee
Note that for $x_1=1$, one finds $k^b_{c,n1}/k^b_{u,n}>1$, which is Eq.~22 in Ref.~\cite{jang-jpcb121}, the agonist criterion for the single odorant case.
The $x_1$-dependency of $A_n$ under this high basal activity condition is well represented in Fig.~\ref{An3d}(c-5), in which the crossover occurs at the threshold fraction. 
The latter is 
\be\label{eq:x_1ast_highbasal}
x_{n1}^{\ast, \text{high}} = \frac{1}{1-\frac{ {K}_{d,n1} (\tilde{k}^f_{c,n1} - \tilde{K}_{c,n1}) }{ {K}_{d,n2} (\tilde{k}^f_{c,n2} - \tilde{K}_{c,n2}) }},
\ee
shown in Fig.~\ref{fig:x1} as a function of ${ ({k}^f_{c,n1}/{k}^b_{u,n}-{K}_{c,n1} ) }/{ ({k}^f_{c,n2}/{k}^b_{u,n}-{K}_{c,n2} ) }$, which is positive for negative $x$-axis values only, meaning a mixture of agonist and inverse agonist.\\

\subsection{Net odor activity depending on $K_{d,nq}$ and $K_{c,nq}$}

Here, we present the odor activity $A_n$ given in Eq.~\ref{eq:An_dimless}, now as a function of association constants, $K_{c,n1}/K_{c,n2}$ and $K_{d,n1}/K_{d,n2}$. 

\begin{figure*}
\includegraphics[width=0.99\textwidth]{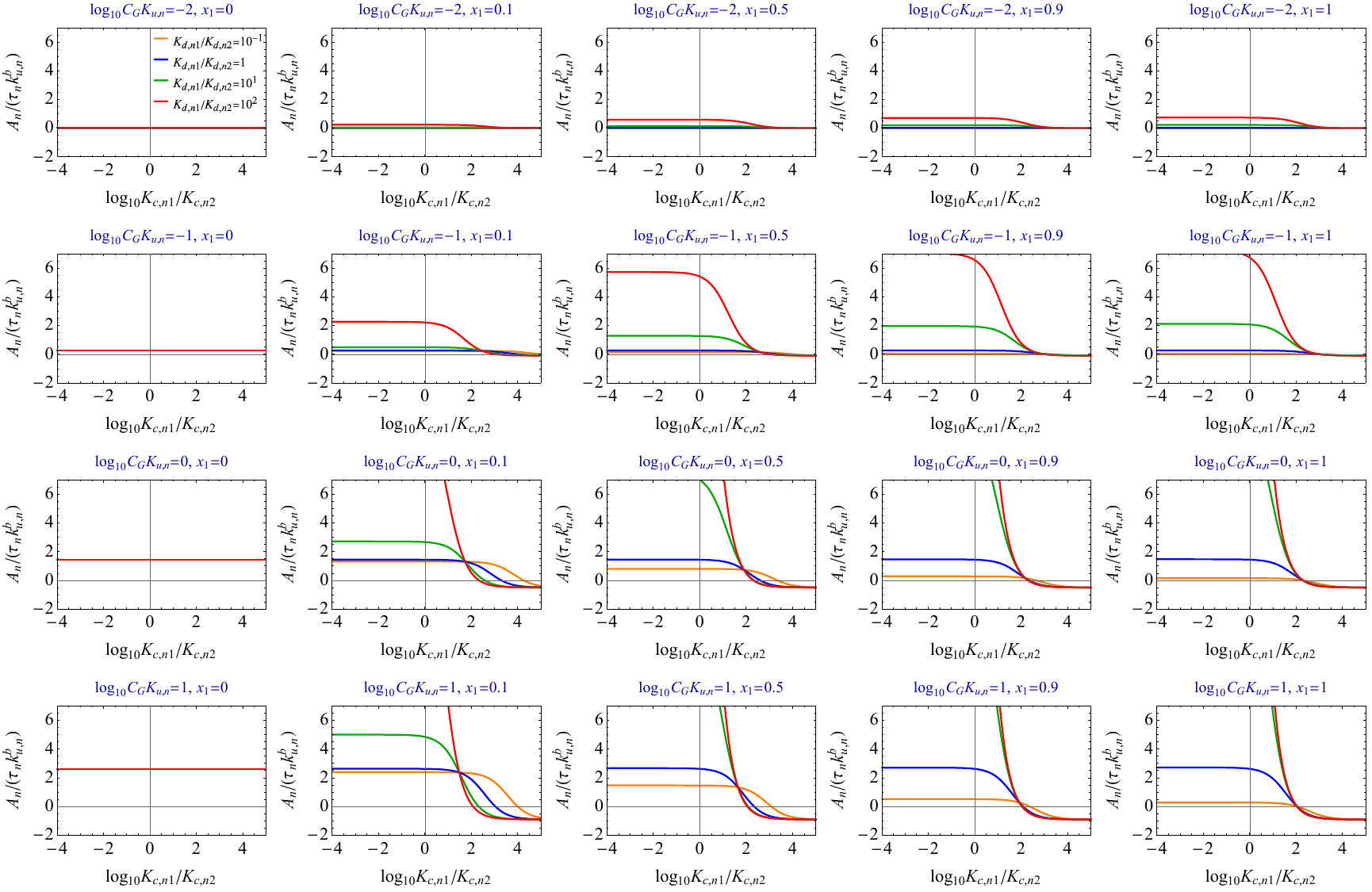}
\caption{\label{fig:SI_An_Kcn} Odor activity $\tilde{A}_n$ in Eq.~\ref{eq:An_dimless} as a function of $K_{c,n1}/K_{c,n2}$ at different values of $C_G K_{u,n}$, $x_1$ and $K_{d,n1}/K_{d,n2}$. We use $\tilde{k}^f_{c,n1}=\tilde{k}^f_{c,n2}=100$, $\tilde{K}_{c,n2}=1$, and $\tilde{K}_{d,n2}=0.03$, following the parameters considered previously~\cite{jang-jpcb121}.}
\end{figure*}

\begin{figure*}
\includegraphics[width=0.99\textwidth]{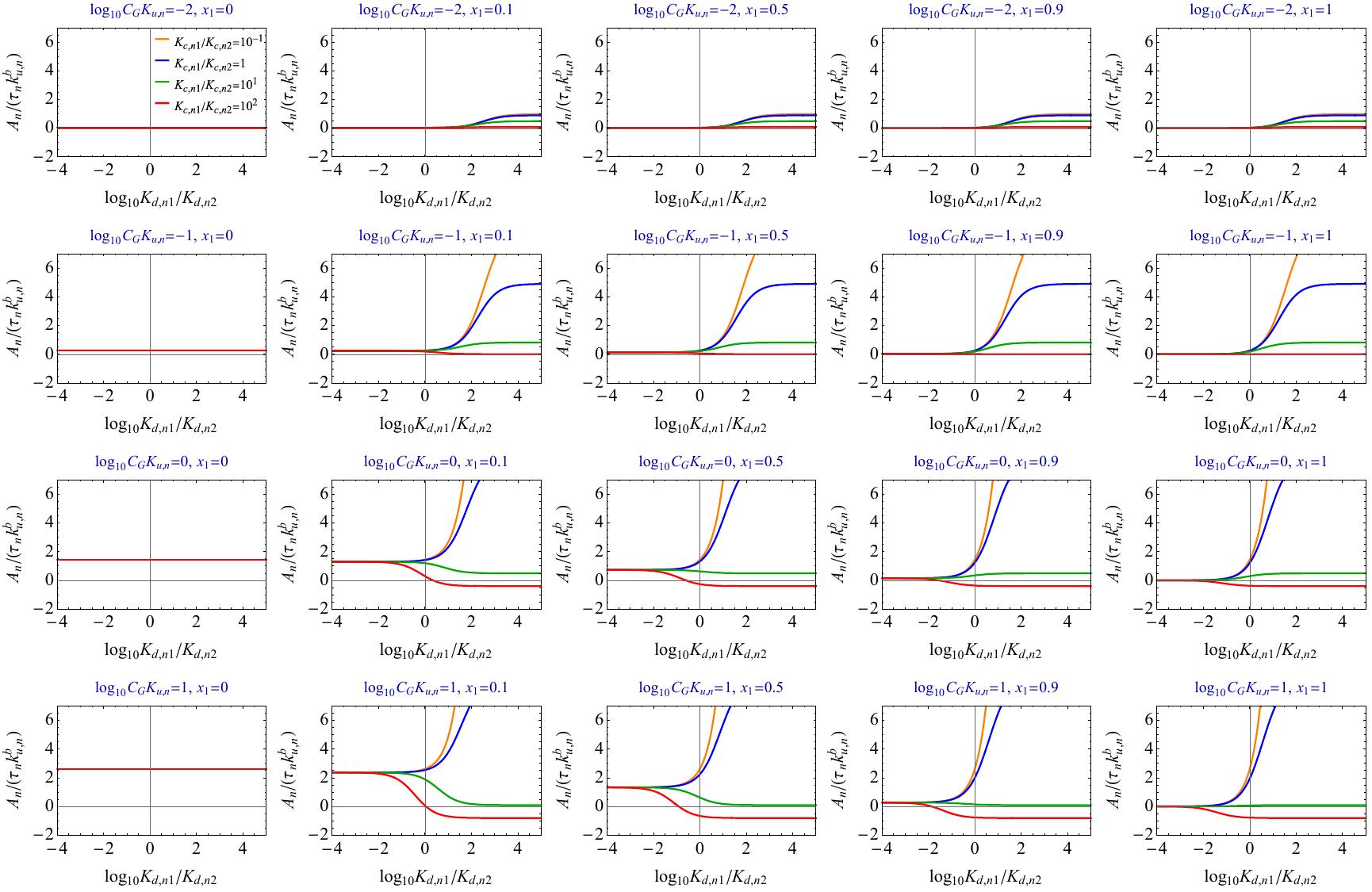}
\caption{\label{fig:SI_An_Kdn} Odor activity $\tilde{A}_n$ in Eq.~\ref{eq:An_dimless} as a function of $K_{d,n1}/K_{d,n2}$ at different values of $C_G K_{u,n}$, $x_1$ and $K_{c,n1}/K_{c,n2}$. We use $\tilde{k}^f_{c,n1}=\tilde{k}^f_{c,n2}=100$, $\tilde{K}_{c,n2}=1$, and $\tilde{K}_{d,n2}=0.03$, following the parameters considered previously~\cite{jang-jpcb121}.}
\end{figure*}

Figure~\ref{fig:SI_An_Kcn} depicts the net odor activity  $A_n(K_{c,n1}/K_{c,n2})$, which provides an overall profile of activity depending on the G-protein association constants.
We also depict the net odor activity  $A_n(K_{d,n1}/K_{d,n2})$ in Fig.~\ref{fig:SI_An_Kdn}, which shows an overall profile of activity depending on the odorant association constants.\\

\begin{thebibliography}{47}%
\makeatletter
\providecommand \@ifxundefined [1]{%
 \@ifx{#1\undefined}
}%
\providecommand \@ifnum [1]{%
 \ifnum #1\expandafter \@firstoftwo
 \else \expandafter \@secondoftwo
 \fi
}%
\providecommand \@ifx [1]{%
 \ifx #1\expandafter \@firstoftwo
 \else \expandafter \@secondoftwo
 \fi
}%
\providecommand \natexlab [1]{#1}%
\providecommand \enquote  [1]{``#1''}%
\providecommand \bibnamefont  [1]{#1}%
\providecommand \bibfnamefont [1]{#1}%
\providecommand \citenamefont [1]{#1}%
\providecommand \href@noop [0]{\@secondoftwo}%
\providecommand \href [0]{\begingroup \@sanitize@url \@href}%
\providecommand \@href[1]{\@@startlink{#1}\@@href}%
\providecommand \@@href[1]{\endgroup#1\@@endlink}%
\providecommand \@sanitize@url [0]{\catcode `\\12\catcode `\$12\catcode
  `\&12\catcode `\#12\catcode `\^12\catcode `\_12\catcode `\%12\relax}%
\providecommand \@@startlink[1]{}%
\providecommand \@@endlink[0]{}%
\providecommand \url  [0]{\begingroup\@sanitize@url \@url }%
\providecommand \@url [1]{\endgroup\@href {#1}{\urlprefix }}%
\providecommand \urlprefix  [0]{URL }%
\providecommand \Eprint [0]{\href }%
\providecommand \doibase [0]{https://doi.org/}%
\providecommand \selectlanguage [0]{\@gobble}%
\providecommand \bibinfo  [0]{\@secondoftwo}%
\providecommand \bibfield  [0]{\@secondoftwo}%
\providecommand \translation [1]{[#1]}%
\providecommand \BibitemOpen [0]{}%
\providecommand \bibitemStop [0]{}%
\providecommand \bibitemNoStop [0]{.\EOS\space}%
\providecommand \EOS [0]{\spacefactor3000\relax}%
\providecommand \BibitemShut  [1]{\csname bibitem#1\endcsname}%
\let\auto@bib@innerbib\@empty
\bibitem [{\citenamefont {Zarzo}(2007)}]{zazro-brcpc82}%
  \BibitemOpen
  \bibfield  {author} {\bibinfo {author} {\bibfnamefont {M.}~\bibnamefont
  {Zarzo}},\ }\bibfield  {title} {\bibinfo {title} {The sense of smell:
  molecular basis of odorant recognition},\ }\href@noop {} {\bibfield
  {journal} {\bibinfo  {journal} {Biol. Rev.}\ }\textbf {\bibinfo {volume}
  {82}},\ \bibinfo {pages} {455} (\bibinfo {year} {2007})}\BibitemShut
  {NoStop}%
\bibitem [{\citenamefont {DeMaria}\ and\ \citenamefont
  {Ngai}(2010)}]{demaria-jcb191}%
  \BibitemOpen
  \bibfield  {author} {\bibinfo {author} {\bibfnamefont {S.}~\bibnamefont
  {DeMaria}}\ and\ \bibinfo {author} {\bibfnamefont {J.}~\bibnamefont {Ngai}},\
  }\bibfield  {title} {\bibinfo {title} {The cell biology of smell},\
  }\href@noop {} {\bibfield  {journal} {\bibinfo  {journal} {J. Cell Biol.}\
  }\textbf {\bibinfo {volume} {191}},\ \bibinfo {pages} {443} (\bibinfo {year}
  {2010})}\BibitemShut {NoStop}%
\bibitem [{\citenamefont {Mori}\ and\ \citenamefont
  {Sakano}(2011)}]{Mori-arn34}%
  \BibitemOpen
  \bibfield  {author} {\bibinfo {author} {\bibfnamefont {K.}~\bibnamefont
  {Mori}}\ and\ \bibinfo {author} {\bibfnamefont {H.}~\bibnamefont {Sakano}},\
  }\bibfield  {title} {\bibinfo {title} {How is the olfactory map formed and
  interpreted in the mammalian brain?},\ }\href@noop {} {\bibfield  {journal}
  {\bibinfo  {journal} {Annu. Rev. Neurosci.}\ }\textbf {\bibinfo {volume}
  {34}},\ \bibinfo {pages} {467} (\bibinfo {year} {2011})}\BibitemShut
  {NoStop}%
\bibitem [{\citenamefont {Tromelin}(2016)}]{tromelin-ffj31}%
  \BibitemOpen
  \bibfield  {author} {\bibinfo {author} {\bibfnamefont {A.}~\bibnamefont
  {Tromelin}},\ }\bibfield  {title} {\bibinfo {title} {Odour perception: A
  review of an intricate signalling pathway},\ }\href@noop {} {\bibfield
  {journal} {\bibinfo  {journal} {Flavour Fragr. J.}\ }\textbf {\bibinfo
  {volume} {31}},\ \bibinfo {pages} {107} (\bibinfo {year} {2016})}\BibitemShut
  {NoStop}%
\bibitem [{\citenamefont {Block}(2018)}]{block-jafc66}%
  \BibitemOpen
  \bibfield  {author} {\bibinfo {author} {\bibfnamefont {E.}~\bibnamefont
  {Block}},\ }\bibfield  {title} {\bibinfo {title} {Molecular basis of
  mammalian odor discrimination: A status report},\ }\href@noop {} {\bibfield
  {journal} {\bibinfo  {journal} {J. Agric. Food Chem.}\ }\textbf {\bibinfo
  {volume} {66}},\ \bibinfo {pages} {13346} (\bibinfo {year}
  {2018})}\BibitemShut {NoStop}%
\bibitem [{\citenamefont {Xu}\ \emph {et~al.}(2023{\natexlab{a}})\citenamefont
  {Xu}, \citenamefont {Zou},\ and\ \citenamefont {Firestein}}]{Xu-fee11}%
  \BibitemOpen
  \bibfield  {author} {\bibinfo {author} {\bibfnamefont {L.}~\bibnamefont
  {Xu}}, \bibinfo {author} {\bibfnamefont {D.-J.}\ \bibnamefont {Zou}},\ and\
  \bibinfo {author} {\bibfnamefont {S.}~\bibnamefont {Firestein}},\ }\bibfield
  {title} {\bibinfo {title} {Odor mixtures: A chord with silent notes},\
  }\href@noop {} {\bibfield  {journal} {\bibinfo  {journal} {Front. Ecol.
  Evol.}\ }\textbf {\bibinfo {volume} {11}},\ \bibinfo {pages} {1135486}
  (\bibinfo {year} {2023}{\natexlab{a}})}\BibitemShut {NoStop}%
\bibitem [{\citenamefont {Bhandawat}\ \emph {et~al.}(2005)\citenamefont
  {Bhandawat}, \citenamefont {Reisert},\ and\ \citenamefont
  {Yau}}]{bhandawat-science308}%
  \BibitemOpen
  \bibfield  {author} {\bibinfo {author} {\bibfnamefont {V.}~\bibnamefont
  {Bhandawat}}, \bibinfo {author} {\bibfnamefont {J.}~\bibnamefont {Reisert}},\
  and\ \bibinfo {author} {\bibfnamefont {K.-W.}\ \bibnamefont {Yau}},\
  }\bibfield  {title} {\bibinfo {title} {Elementary response of olfactory
  receptor neurons to odorants},\ }\href@noop {} {\bibfield  {journal}
  {\bibinfo  {journal} {Science}\ }\textbf {\bibinfo {volume} {308}},\ \bibinfo
  {pages} {1931} (\bibinfo {year} {2005})}\BibitemShut {NoStop}%
\bibitem [{\citenamefont {Bhandawat}\ \emph {et~al.}(2010)\citenamefont
  {Bhandawat}, \citenamefont {Reisert},\ and\ \citenamefont
  {Yau}}]{bhandawat-pnas107}%
  \BibitemOpen
  \bibfield  {author} {\bibinfo {author} {\bibfnamefont {V.}~\bibnamefont
  {Bhandawat}}, \bibinfo {author} {\bibfnamefont {J.}~\bibnamefont {Reisert}},\
  and\ \bibinfo {author} {\bibfnamefont {K.-W.}\ \bibnamefont {Yau}},\
  }\bibfield  {title} {\bibinfo {title} {Signaling by olfactory receptor
  neurons near threshold},\ }\href@noop {} {\bibfield  {journal} {\bibinfo
  {journal} {Proc. Natl. Acad. Sci. U.S.A.}\ }\textbf {\bibinfo {volume}
  {107}},\ \bibinfo {pages} {18682} (\bibinfo {year} {2010})}\BibitemShut
  {NoStop}%
\bibitem [{\citenamefont {Ben-Chaim}\ \emph {et~al.}(2011)\citenamefont
  {Ben-Chaim}, \citenamefont {Cheng},\ and\ \citenamefont
  {Yau}}]{ben-chaim-pnas108}%
  \BibitemOpen
  \bibfield  {author} {\bibinfo {author} {\bibfnamefont {Y.}~\bibnamefont
  {Ben-Chaim}}, \bibinfo {author} {\bibfnamefont {M.~M.}\ \bibnamefont
  {Cheng}},\ and\ \bibinfo {author} {\bibfnamefont {K.-W.}\ \bibnamefont
  {Yau}},\ }\bibfield  {title} {\bibinfo {title} {Unitary response of mouse
  olfactory receptor neurons},\ }\href@noop {} {\bibfield  {journal} {\bibinfo
  {journal} {Proc. Natl. Acad. Sci. U.S.A.}\ }\textbf {\bibinfo {volume}
  {108}},\ \bibinfo {pages} {822} (\bibinfo {year} {2011})}\BibitemShut
  {NoStop}%
\bibitem [{\citenamefont {Buck}\ and\ \citenamefont
  {Axel}(1991)}]{buck-cell65}%
  \BibitemOpen
  \bibfield  {author} {\bibinfo {author} {\bibfnamefont {L.}~\bibnamefont
  {Buck}}\ and\ \bibinfo {author} {\bibfnamefont {R.}~\bibnamefont {Axel}},\
  }\bibfield  {title} {\bibinfo {title} {A novel multigene family may encode
  odorant receptors: a molecular basis for odor recognition},\ }\href@noop {}
  {\bibfield  {journal} {\bibinfo  {journal} {Cell}\ }\textbf {\bibinfo
  {volume} {65}},\ \bibinfo {pages} {175} (\bibinfo {year} {1991})}\BibitemShut
  {NoStop}%
\bibitem [{\citenamefont {Li}\ \emph {et~al.}(2014)\citenamefont {Li},
  \citenamefont {Peterlin}, \citenamefont {Ho}, \citenamefont {Yarnitzky},
  \citenamefont {Liu}, \citenamefont {Fichman}, \citenamefont {Niv},
  \citenamefont {Matsunami}, \citenamefont {Firestein},\ and\ \citenamefont
  {Ryan}}]{Li-acb9}%
  \BibitemOpen
  \bibfield  {author} {\bibinfo {author} {\bibfnamefont {Y.}~\bibnamefont
  {Li}}, \bibinfo {author} {\bibfnamefont {Z.}~\bibnamefont {Peterlin}},
  \bibinfo {author} {\bibfnamefont {J.}~\bibnamefont {Ho}}, \bibinfo {author}
  {\bibfnamefont {T.}~\bibnamefont {Yarnitzky}}, \bibinfo {author}
  {\bibfnamefont {M.~T.}\ \bibnamefont {Liu}}, \bibinfo {author} {\bibfnamefont
  {M.}~\bibnamefont {Fichman}}, \bibinfo {author} {\bibfnamefont {M.~Y.}\
  \bibnamefont {Niv}}, \bibinfo {author} {\bibfnamefont {H.}~\bibnamefont
  {Matsunami}}, \bibinfo {author} {\bibfnamefont {S.}~\bibnamefont
  {Firestein}},\ and\ \bibinfo {author} {\bibfnamefont {K.}~\bibnamefont
  {Ryan}},\ }\bibfield  {title} {\bibinfo {title} {Aldehyde recognition and
  discrimination by mammalian odorant receptors via functional group-specific
  hydration chemistry},\ }\href@noop {} {\bibfield  {journal} {\bibinfo
  {journal} {ACS Chem. Biol.}\ }\textbf {\bibinfo {volume} {9}},\ \bibinfo
  {pages} {2563} (\bibinfo {year} {2014})}\BibitemShut {NoStop}%
\bibitem [{\citenamefont {Stanislav A.~Pshenichnyuk}\ \emph
  {et~al.}(2018)\citenamefont {Stanislav A.~Pshenichnyuk}, \citenamefont
  {Rakhmeyev}, \citenamefont {Asfandiarov}, \citenamefont {Komolov},
  \citenamefont {Modelli},\ and\ \citenamefont {Jones}}]{pshenichnyuk-jpcl9}%
  \BibitemOpen
  \bibfield  {author} {\bibinfo {author} {\bibfnamefont {S.~A.}\ \bibnamefont
  {Stanislav A.~Pshenichnyuk}}, \bibinfo {author} {\bibfnamefont {R.~G.}\
  \bibnamefont {Rakhmeyev}}, \bibinfo {author} {\bibfnamefont {N.~L.}\
  \bibnamefont {Asfandiarov}}, \bibinfo {author} {\bibfnamefont {A.~S.}\
  \bibnamefont {Komolov}}, \bibinfo {author} {\bibfnamefont {A.}~\bibnamefont
  {Modelli}},\ and\ \bibinfo {author} {\bibfnamefont {D.}~\bibnamefont
  {Jones}},\ }\bibfield  {title} {\bibinfo {title} {Can the electron-accepting
  properties of odorants be involved in their recognition by the olfactory
  system?},\ }\href@noop {} {\bibfield  {journal} {\bibinfo  {journal} {J.
  Phys. Chem. Lett.}\ }\textbf {\bibinfo {volume} {9}},\ \bibinfo {pages}
  {2320} (\bibinfo {year} {2018})}\BibitemShut {NoStop}%
\bibitem [{\citenamefont {Jang}\ and\ \citenamefont
  {Hyeon}(2017)}]{jang-jpcb121}%
  \BibitemOpen
  \bibfield  {author} {\bibinfo {author} {\bibfnamefont {S.}~\bibnamefont
  {Jang}}\ and\ \bibinfo {author} {\bibfnamefont {C.}~\bibnamefont {Hyeon}},\
  }\bibfield  {title} {\bibinfo {title} {Kinetic model for the activation of
  mammalian olfactory receptor},\ }\href@noop {} {\bibfield  {journal}
  {\bibinfo  {journal} {J. Phys. Chem. B}\ }\textbf {\bibinfo {volume} {121}},\
  \bibinfo {pages} {1304} (\bibinfo {year} {2017})}\BibitemShut {NoStop}%
\bibitem [{\citenamefont {Bushdid}\ \emph {et~al.}(2018)\citenamefont
  {Bushdid}, \citenamefont {de~March}, \citenamefont {Fiorucci}, \citenamefont
  {Matsunami},\ and\ \citenamefont {Golebiowski}}]{bushdid-jpcl9}%
  \BibitemOpen
  \bibfield  {author} {\bibinfo {author} {\bibfnamefont {C.}~\bibnamefont
  {Bushdid}}, \bibinfo {author} {\bibfnamefont {C.~A.}\ \bibnamefont
  {de~March}}, \bibinfo {author} {\bibfnamefont {S.}~\bibnamefont {Fiorucci}},
  \bibinfo {author} {\bibfnamefont {H.}~\bibnamefont {Matsunami}},\ and\
  \bibinfo {author} {\bibfnamefont {J.}~\bibnamefont {Golebiowski}},\
  }\bibfield  {title} {\bibinfo {title} {Agonists of g-protein-coupled odorant
  receptors are predicted from chemical features},\ }\href@noop {} {\bibfield
  {journal} {\bibinfo  {journal} {J. Phys. Chem. Lett.}\ }\textbf {\bibinfo
  {volume} {9}},\ \bibinfo {pages} {2235} (\bibinfo {year} {2018})}\BibitemShut
  {NoStop}%
\bibitem [{\citenamefont {de~March}\ \emph {et~al.}(2015)\citenamefont
  {de~March}, \citenamefont {Kim}, \citenamefont {Antonczak}, \citenamefont
  {Goddard~III},\ and\ \citenamefont {Golebiowski}}]{demarch-ps24}%
  \BibitemOpen
  \bibfield  {author} {\bibinfo {author} {\bibfnamefont {C.~A.}\ \bibnamefont
  {de~March}}, \bibinfo {author} {\bibfnamefont {S.-K.}\ \bibnamefont {Kim}},
  \bibinfo {author} {\bibfnamefont {S.}~\bibnamefont {Antonczak}}, \bibinfo
  {author} {\bibfnamefont {W.~A.}\ \bibnamefont {Goddard~III}},\ and\ \bibinfo
  {author} {\bibfnamefont {J.}~\bibnamefont {Golebiowski}},\ }\bibfield
  {title} {\bibinfo {title} {G protein-coupled odorant receptors: From sequence
  to structure},\ }\href@noop {} {\bibfield  {journal} {\bibinfo  {journal}
  {Protein Sci.}\ }\textbf {\bibinfo {volume} {24}},\ \bibinfo {pages} {1543}
  (\bibinfo {year} {2015})}\BibitemShut {NoStop}%
\bibitem [{\citenamefont {Huang}\ \emph {et~al.}(2015)\citenamefont {Huang},
  \citenamefont {Manglik}, \citenamefont {Venkatakrishnan}, \citenamefont
  {Laeremans}, \citenamefont {Feinberg}, \citenamefont {Sanborn}, \citenamefont
  {Kato}, \citenamefont {Livingston}, \citenamefont {Thorsen}, \citenamefont
  {Kling} \emph {et~al.}}]{huang-nature524}%
  \BibitemOpen
  \bibfield  {author} {\bibinfo {author} {\bibfnamefont {W.}~\bibnamefont
  {Huang}}, \bibinfo {author} {\bibfnamefont {A.}~\bibnamefont {Manglik}},
  \bibinfo {author} {\bibfnamefont {A.}~\bibnamefont {Venkatakrishnan}},
  \bibinfo {author} {\bibfnamefont {T.}~\bibnamefont {Laeremans}}, \bibinfo
  {author} {\bibfnamefont {E.~N.}\ \bibnamefont {Feinberg}}, \bibinfo {author}
  {\bibfnamefont {A.~L.}\ \bibnamefont {Sanborn}}, \bibinfo {author}
  {\bibfnamefont {H.~E.}\ \bibnamefont {Kato}}, \bibinfo {author}
  {\bibfnamefont {K.~E.}\ \bibnamefont {Livingston}}, \bibinfo {author}
  {\bibfnamefont {T.~S.}\ \bibnamefont {Thorsen}}, \bibinfo {author}
  {\bibfnamefont {R.~C.}\ \bibnamefont {Kling}}, \emph {et~al.},\ }\bibfield
  {title} {\bibinfo {title} {Structural insights into $\mu$-opioid receptor
  activation},\ }\href@noop {} {\bibfield  {journal} {\bibinfo  {journal}
  {Nature}\ }\textbf {\bibinfo {volume} {524}},\ \bibinfo {pages} {315}
  (\bibinfo {year} {2015})}\BibitemShut {NoStop}%
\bibitem [{\citenamefont {Sounier}\ \emph {et~al.}(2015)\citenamefont
  {Sounier}, \citenamefont {Mas}, \citenamefont {Steyaert}, \citenamefont
  {Laeremans}, \citenamefont {Manglik}, \citenamefont {Huang}, \citenamefont
  {Kobilka}, \citenamefont {D{\'e}m{\'e}n{\'e}},\ and\ \citenamefont
  {Granier}}]{sounier-nature524}%
  \BibitemOpen
  \bibfield  {author} {\bibinfo {author} {\bibfnamefont {R.}~\bibnamefont
  {Sounier}}, \bibinfo {author} {\bibfnamefont {C.}~\bibnamefont {Mas}},
  \bibinfo {author} {\bibfnamefont {J.}~\bibnamefont {Steyaert}}, \bibinfo
  {author} {\bibfnamefont {T.}~\bibnamefont {Laeremans}}, \bibinfo {author}
  {\bibfnamefont {A.}~\bibnamefont {Manglik}}, \bibinfo {author} {\bibfnamefont
  {W.}~\bibnamefont {Huang}}, \bibinfo {author} {\bibfnamefont {B.~K.}\
  \bibnamefont {Kobilka}}, \bibinfo {author} {\bibfnamefont {H.}~\bibnamefont
  {D{\'e}m{\'e}n{\'e}}},\ and\ \bibinfo {author} {\bibfnamefont
  {S.}~\bibnamefont {Granier}},\ }\bibfield  {title} {\bibinfo {title}
  {Propagation of conformational changes during $\mu$-opioid receptor
  activation},\ }\href@noop {} {\bibfield  {journal} {\bibinfo  {journal}
  {Nature}\ }\textbf {\bibinfo {volume} {524}},\ \bibinfo {pages} {375}
  (\bibinfo {year} {2015})}\BibitemShut {NoStop}%
\bibitem [{\citenamefont {Manglik}\ \emph {et~al.}(2015)\citenamefont
  {Manglik}, \citenamefont {Kim}, \citenamefont {Masureel}, \citenamefont
  {Altenbach}, \citenamefont {Yang}, \citenamefont {Hilger}, \citenamefont
  {Lerch}, \citenamefont {Kobilka}, \citenamefont {Thian}, \citenamefont
  {Hubbell} \emph {et~al.}}]{manglik-cell161}%
  \BibitemOpen
  \bibfield  {author} {\bibinfo {author} {\bibfnamefont {A.}~\bibnamefont
  {Manglik}}, \bibinfo {author} {\bibfnamefont {T.~H.}\ \bibnamefont {Kim}},
  \bibinfo {author} {\bibfnamefont {M.}~\bibnamefont {Masureel}}, \bibinfo
  {author} {\bibfnamefont {C.}~\bibnamefont {Altenbach}}, \bibinfo {author}
  {\bibfnamefont {Z.}~\bibnamefont {Yang}}, \bibinfo {author} {\bibfnamefont
  {D.}~\bibnamefont {Hilger}}, \bibinfo {author} {\bibfnamefont {M.~T.}\
  \bibnamefont {Lerch}}, \bibinfo {author} {\bibfnamefont {T.~S.}\ \bibnamefont
  {Kobilka}}, \bibinfo {author} {\bibfnamefont {F.~S.}\ \bibnamefont {Thian}},
  \bibinfo {author} {\bibfnamefont {W.~L.}\ \bibnamefont {Hubbell}}, \emph
  {et~al.},\ }\bibfield  {title} {\bibinfo {title} {Structural insights into
  the dynamic process of $\beta$2-adrenergic receptor signaling},\ }\href@noop
  {} {\bibfield  {journal} {\bibinfo  {journal} {Cell}\ }\textbf {\bibinfo
  {volume} {161}},\ \bibinfo {pages} {1101} (\bibinfo {year}
  {2015})}\BibitemShut {NoStop}%
\bibitem [{\citenamefont {Dror}\ \emph {et~al.}(2011)\citenamefont {Dror},
  \citenamefont {Pan}, \citenamefont {Arlow}, \citenamefont {Borhani},
  \citenamefont {Maragakis}, \citenamefont {Shan}, \citenamefont {Xu},\ and\
  \citenamefont {Shaw}}]{dror-pnas108}%
  \BibitemOpen
  \bibfield  {author} {\bibinfo {author} {\bibfnamefont {R.~O.}\ \bibnamefont
  {Dror}}, \bibinfo {author} {\bibfnamefont {A.~C.}\ \bibnamefont {Pan}},
  \bibinfo {author} {\bibfnamefont {D.~H.}\ \bibnamefont {Arlow}}, \bibinfo
  {author} {\bibfnamefont {D.~W.}\ \bibnamefont {Borhani}}, \bibinfo {author}
  {\bibfnamefont {P.}~\bibnamefont {Maragakis}}, \bibinfo {author}
  {\bibfnamefont {Y.}~\bibnamefont {Shan}}, \bibinfo {author} {\bibfnamefont
  {H.}~\bibnamefont {Xu}},\ and\ \bibinfo {author} {\bibfnamefont {D.~E.}\
  \bibnamefont {Shaw}},\ }\bibfield  {title} {\bibinfo {title} {Pathway and
  mechanism of drug binding to g-protein-coupled receptors},\ }\href@noop {}
  {\bibfield  {journal} {\bibinfo  {journal} {Proc. Natl. Acad. Sci., U.S.A.}\
  }\textbf {\bibinfo {volume} {108}},\ \bibinfo {pages} {13118} (\bibinfo
  {year} {2011})}\BibitemShut {NoStop}%
\bibitem [{\citenamefont {Lee}\ \emph {et~al.}(2016)\citenamefont {Lee},
  \citenamefont {Kim}, \citenamefont {Choi},\ and\ \citenamefont
  {Hyeon}}]{lee-bj111}%
  \BibitemOpen
  \bibfield  {author} {\bibinfo {author} {\bibfnamefont {Y.}~\bibnamefont
  {Lee}}, \bibinfo {author} {\bibfnamefont {S.}~\bibnamefont {Kim}}, \bibinfo
  {author} {\bibfnamefont {S.}~\bibnamefont {Choi}},\ and\ \bibinfo {author}
  {\bibfnamefont {C.}~\bibnamefont {Hyeon}},\ }\bibfield  {title} {\bibinfo
  {title} {Ultraslow water-mediated transmembrane interactions regulate the
  activation of a2a adenosine receptor},\ }\href@noop {} {\bibfield  {journal}
  {\bibinfo  {journal} {Biophys. J.}\ }\textbf {\bibinfo {volume} {111}},\
  \bibinfo {pages} {1180} (\bibinfo {year} {2016})}\BibitemShut {NoStop}%
\bibitem [{\citenamefont {Lee}\ \emph {et~al.}(2015)\citenamefont {Lee},
  \citenamefont {Choi},\ and\ \citenamefont {Hyeon}}]{lee-pcb11}%
  \BibitemOpen
  \bibfield  {author} {\bibinfo {author} {\bibfnamefont {Y.}~\bibnamefont
  {Lee}}, \bibinfo {author} {\bibfnamefont {S.}~\bibnamefont {Choi}},\ and\
  \bibinfo {author} {\bibfnamefont {C.}~\bibnamefont {Hyeon}},\ }\bibfield
  {title} {\bibinfo {title} {Communication over the network of binary switches
  regulates the activation of a2a adenosine receptor},\ }\href@noop {}
  {\bibfield  {journal} {\bibinfo  {journal} {PLOS Comp. Biol.}\ }\textbf
  {\bibinfo {volume} {11}},\ \bibinfo {pages} {e1004044} (\bibinfo {year}
  {2015})}\BibitemShut {NoStop}%
\bibitem [{\citenamefont {Cornish-Bowden}(2015)}]{cornish-bowden-ps4}%
  \BibitemOpen
  \bibfield  {author} {\bibinfo {author} {\bibnamefont {Cornish-Bowden}},\
  }\bibfield  {title} {\bibinfo {title} {One hundred years of michaelis-menten
  kinetics},\ }\href@noop {} {\bibfield  {journal} {\bibinfo  {journal}
  {Perspect. Sci.}\ }\textbf {\bibinfo {volume} {4}},\ \bibinfo {pages} {3}
  (\bibinfo {year} {2015})}\BibitemShut {NoStop}%
\bibitem [{Note2()}]{Note2}%
  \BibitemOpen
  \bibinfo {note} {Defined as the concentration of the odorant where the signal
  strength becomes half the maximum.}\BibitemShut {Stop}%
\bibitem [{\citenamefont {Cruz}\ and\ \citenamefont
  {Lowe}(2013)}]{cruz-scirep3}%
  \BibitemOpen
  \bibfield  {author} {\bibinfo {author} {\bibfnamefont {G.}~\bibnamefont
  {Cruz}}\ and\ \bibinfo {author} {\bibfnamefont {G.}~\bibnamefont {Lowe}},\
  }\bibfield  {title} {\bibinfo {title} {Neural coding of binary mixtures in a
  structurally related odorant pair},\ }\href@noop {} {\bibfield  {journal}
  {\bibinfo  {journal} {Sci. Rep.}\ }\textbf {\bibinfo {volume} {3}},\ \bibinfo
  {pages} {1220} (\bibinfo {year} {2013})}\BibitemShut {NoStop}%
\bibitem [{\citenamefont {Reddy}\ \emph
  {et~al.}(2018{\natexlab{a}})\citenamefont {Reddy}, \citenamefont {Zak},
  \citenamefont {Vergassola},\ and\ \citenamefont {Murthy}}]{reddy-elife7}%
  \BibitemOpen
  \bibfield  {author} {\bibinfo {author} {\bibfnamefont {G.}~\bibnamefont
  {Reddy}}, \bibinfo {author} {\bibfnamefont {J.~D.}\ \bibnamefont {Zak}},
  \bibinfo {author} {\bibfnamefont {M.}~\bibnamefont {Vergassola}},\ and\
  \bibinfo {author} {\bibfnamefont {V.~N.}\ \bibnamefont {Murthy}},\ }\bibfield
   {title} {\bibinfo {title} {Antagonism in olfactory receptor neurons and its
  implications for the perception of odor mixtures},\ }\href@noop {} {\bibfield
   {journal} {\bibinfo  {journal} {eLife}\ }\textbf {\bibinfo {volume} {7}},\
  \bibinfo {pages} {e34958} (\bibinfo {year} {2018}{\natexlab{a}})}\BibitemShut
  {NoStop}%
\bibitem [{\citenamefont {Singh}\ \emph {et~al.}(2019)\citenamefont {Singh},
  \citenamefont {Murphy}, \citenamefont {Balasubramanian},\ and\ \citenamefont
  {Mainland}}]{singh-pnas116}%
  \BibitemOpen
  \bibfield  {author} {\bibinfo {author} {\bibfnamefont {V.}~\bibnamefont
  {Singh}}, \bibinfo {author} {\bibfnamefont {N.~R.}\ \bibnamefont {Murphy}},
  \bibinfo {author} {\bibfnamefont {V.}~\bibnamefont {Balasubramanian}},\ and\
  \bibinfo {author} {\bibfnamefont {J.~D.}\ \bibnamefont {Mainland}},\
  }\bibfield  {title} {\bibinfo {title} {Competitive binding predicts nonlinear
  responses of olfactory receptors to complex mixtures},\ }\href@noop {}
  {\bibfield  {journal} {\bibinfo  {journal} {Proc. Natl. Acad. Sci., U.S.A.}\
  }\textbf {\bibinfo {volume} {116}},\ \bibinfo {pages} {9598} (\bibinfo {year}
  {2019})}\BibitemShut {NoStop}%
\bibitem [{\citenamefont {Xu}\ \emph {et~al.}(2020)\citenamefont {Xu},
  \citenamefont {Li}, \citenamefont {Voleti}, \citenamefont {Zou},
  \citenamefont {Hillman},\ and\ \citenamefont {Firestein}}]{xu-science368}%
  \BibitemOpen
  \bibfield  {author} {\bibinfo {author} {\bibfnamefont {L.}~\bibnamefont
  {Xu}}, \bibinfo {author} {\bibfnamefont {W.}~\bibnamefont {Li}}, \bibinfo
  {author} {\bibfnamefont {V.}~\bibnamefont {Voleti}}, \bibinfo {author}
  {\bibfnamefont {D.-J.}\ \bibnamefont {Zou}}, \bibinfo {author} {\bibfnamefont
  {E.~M.}\ \bibnamefont {Hillman}},\ and\ \bibinfo {author} {\bibfnamefont
  {S.}~\bibnamefont {Firestein}},\ }\bibfield  {title} {\bibinfo {title}
  {Widespread receptor-driven modulation in peripheral olfactory coding},\
  }\href@noop {} {\bibfield  {journal} {\bibinfo  {journal} {Science}\ }\textbf
  {\bibinfo {volume} {368}},\ \bibinfo {pages} {eaaz5390} (\bibinfo {year}
  {2020})}\BibitemShut {NoStop}%
\bibitem [{\citenamefont {Kurian}\ \emph {et~al.}(2021)\citenamefont {Kurian},
  \citenamefont {Naressi}, \citenamefont {Manoel}, \citenamefont {Barwich},
  \citenamefont {Malnic},\ and\ \citenamefont {Saraiva}}]{kurian-ctr383}%
  \BibitemOpen
  \bibfield  {author} {\bibinfo {author} {\bibfnamefont {S.~M.}\ \bibnamefont
  {Kurian}}, \bibinfo {author} {\bibfnamefont {R.~G.}\ \bibnamefont {Naressi}},
  \bibinfo {author} {\bibfnamefont {D.}~\bibnamefont {Manoel}}, \bibinfo
  {author} {\bibfnamefont {A.-S.}\ \bibnamefont {Barwich}}, \bibinfo {author}
  {\bibfnamefont {B.}~\bibnamefont {Malnic}},\ and\ \bibinfo {author}
  {\bibfnamefont {L.~R.}\ \bibnamefont {Saraiva}},\ }\bibfield  {title}
  {\bibinfo {title} {Odor coding in the mammalian olfactory epithelium},\
  }\href@noop {} {\bibfield  {journal} {\bibinfo  {journal} {Cell Tissue Res.}\
  }\textbf {\bibinfo {volume} {383}},\ \bibinfo {pages} {445} (\bibinfo {year}
  {2021})}\BibitemShut {NoStop}%
\bibitem [{\citenamefont {Marasco}\ \emph {et~al.}(2016)\citenamefont
  {Marasco}, \citenamefont {De~Paris},\ and\ \citenamefont
  {Migliore}}]{marasco-scirep6}%
  \BibitemOpen
  \bibfield  {author} {\bibinfo {author} {\bibfnamefont {A.}~\bibnamefont
  {Marasco}}, \bibinfo {author} {\bibfnamefont {A.}~\bibnamefont {De~Paris}},\
  and\ \bibinfo {author} {\bibfnamefont {M.}~\bibnamefont {Migliore}},\
  }\bibfield  {title} {\bibinfo {title} {Predicting the response of olfactory
  sensory neurons to odor mixtures from single odor response},\ }\href@noop {}
  {\bibfield  {journal} {\bibinfo  {journal} {Sci. Rep.}\ }\textbf {\bibinfo
  {volume} {6}},\ \bibinfo {pages} {24091} (\bibinfo {year}
  {2016})}\BibitemShut {NoStop}%
\bibitem [{\citenamefont {Spehr}\ and\ \citenamefont
  {Munger}(2009)}]{spehr-jn109}%
  \BibitemOpen
  \bibfield  {author} {\bibinfo {author} {\bibfnamefont {M.}~\bibnamefont
  {Spehr}}\ and\ \bibinfo {author} {\bibfnamefont {S.~D.}\ \bibnamefont
  {Munger}},\ }\bibfield  {title} {\bibinfo {title} {Olfactory receptors: G
  protein-coupled receptors and beyond},\ }\href@noop {} {\bibfield  {journal}
  {\bibinfo  {journal} {J. Neurochem.}\ }\textbf {\bibinfo {volume} {109}},\
  \bibinfo {pages} {1570} (\bibinfo {year} {2009})}\BibitemShut {NoStop}%
\bibitem [{\citenamefont {Niimura}\ and\ \citenamefont
  {Nei}(2006)}]{niimura-2006}%
  \BibitemOpen
  \bibfield  {author} {\bibinfo {author} {\bibfnamefont {Y.}~\bibnamefont
  {Niimura}}\ and\ \bibinfo {author} {\bibfnamefont {M.}~\bibnamefont {Nei}},\
  }\bibfield  {title} {\bibinfo {title} {Evolutionary dynamics of olfactory and
  other chemosensory receptor genes in vertebrates},\ }\href@noop {} {\bibfield
   {journal} {\bibinfo  {journal} {J. Hum. Genet.}\ }\textbf {\bibinfo {volume}
  {51}},\ \bibinfo {pages} {505} (\bibinfo {year} {2006})}\BibitemShut
  {NoStop}%
\bibitem [{\citenamefont {Billesb{\o}lle}\ \emph {et~al.}(2023)\citenamefont
  {Billesb{\o}lle}, \citenamefont {de~March}, \citenamefont {van~der Velden},
  \citenamefont {Ma}, \citenamefont {Tewari}, \citenamefont {Del~Torrent},
  \citenamefont {Li}, \citenamefont {Faust}, \citenamefont {Vaidehi},
  \citenamefont {Matsunami} \emph {et~al.}}]{billesbolle2023structural}%
  \BibitemOpen
  \bibfield  {author} {\bibinfo {author} {\bibfnamefont {C.~B.}\ \bibnamefont
  {Billesb{\o}lle}}, \bibinfo {author} {\bibfnamefont {C.~A.}\ \bibnamefont
  {de~March}}, \bibinfo {author} {\bibfnamefont {W.~J.}\ \bibnamefont {van~der
  Velden}}, \bibinfo {author} {\bibfnamefont {N.}~\bibnamefont {Ma}}, \bibinfo
  {author} {\bibfnamefont {J.}~\bibnamefont {Tewari}}, \bibinfo {author}
  {\bibfnamefont {C.~L.}\ \bibnamefont {Del~Torrent}}, \bibinfo {author}
  {\bibfnamefont {L.}~\bibnamefont {Li}}, \bibinfo {author} {\bibfnamefont
  {B.}~\bibnamefont {Faust}}, \bibinfo {author} {\bibfnamefont
  {N.}~\bibnamefont {Vaidehi}}, \bibinfo {author} {\bibfnamefont
  {H.}~\bibnamefont {Matsunami}}, \emph {et~al.},\ }\bibfield  {title}
  {\bibinfo {title} {Structural basis of odorant recognition by a human odorant
  receptor},\ }\href@noop {} {\bibfield  {journal} {\bibinfo  {journal}
  {Nature}\ }\textbf {\bibinfo {volume} {615}},\ \bibinfo {pages} {742}
  (\bibinfo {year} {2023})}\BibitemShut {NoStop}%
\bibitem [{\citenamefont {Rosenbaum}\ \emph {et~al.}(2009)\citenamefont
  {Rosenbaum}, \citenamefont {Rasmussen},\ and\ \citenamefont
  {Kobilka}}]{rosenbaum-nature459}%
  \BibitemOpen
  \bibfield  {author} {\bibinfo {author} {\bibfnamefont {D.~M.}\ \bibnamefont
  {Rosenbaum}}, \bibinfo {author} {\bibfnamefont {S.~G.}\ \bibnamefont
  {Rasmussen}},\ and\ \bibinfo {author} {\bibfnamefont {B.~K.}\ \bibnamefont
  {Kobilka}},\ }\bibfield  {title} {\bibinfo {title} {The structure and
  function of g-protein-coupled receptors},\ }\href@noop {} {\bibfield
  {journal} {\bibinfo  {journal} {Nature}\ }\textbf {\bibinfo {volume} {459}},\
  \bibinfo {pages} {356} (\bibinfo {year} {2009})}\BibitemShut {NoStop}%
\bibitem [{\citenamefont {Berbari}\ \emph {et~al.}(2009)\citenamefont
  {Berbari}, \citenamefont {O'Connor}, \citenamefont {Haycraft},\ and\
  \citenamefont {Yoder}}]{berbari-cb19}%
  \BibitemOpen
  \bibfield  {author} {\bibinfo {author} {\bibfnamefont {N.~F.}\ \bibnamefont
  {Berbari}}, \bibinfo {author} {\bibfnamefont {A.~K.}\ \bibnamefont
  {O'Connor}}, \bibinfo {author} {\bibfnamefont {C.~J.}\ \bibnamefont
  {Haycraft}},\ and\ \bibinfo {author} {\bibfnamefont {B.~K.}\ \bibnamefont
  {Yoder}},\ }\bibfield  {title} {\bibinfo {title} {The primary cilium as a
  complex signaling center},\ }\href@noop {} {\bibfield  {journal} {\bibinfo
  {journal} {Curr. Biol.}\ }\textbf {\bibinfo {volume} {19}},\ \bibinfo {pages}
  {R526} (\bibinfo {year} {2009})}\BibitemShut {NoStop}%
\bibitem [{\citenamefont {Inagaki}\ \emph {et~al.}(2020)\citenamefont
  {Inagaki}, \citenamefont {Iwata}, \citenamefont {Iwamoto},\ and\
  \citenamefont {Imai}}]{Inagaki-2020}%
  \BibitemOpen
  \bibfield  {author} {\bibinfo {author} {\bibfnamefont {S.}~\bibnamefont
  {Inagaki}}, \bibinfo {author} {\bibfnamefont {R.}~\bibnamefont {Iwata}},
  \bibinfo {author} {\bibfnamefont {M.}~\bibnamefont {Iwamoto}},\ and\ \bibinfo
  {author} {\bibfnamefont {T.}~\bibnamefont {Imai}},\ }\bibfield  {title}
  {\bibinfo {title} {Widespread inhibition, antagonism, and synergy in mouse
  olfactory sensory neurons in vivo},\ }\href@noop {} {\bibfield  {journal}
  {\bibinfo  {journal} {Cell Rep.}\ }\textbf {\bibinfo {volume} {31}} (\bibinfo
  {year} {2020})}\BibitemShut {NoStop}%
\bibitem [{\citenamefont {Pfister}\ \emph {et~al.}(2020)\citenamefont
  {Pfister}, \citenamefont {Smith}, \citenamefont {Evans}, \citenamefont
  {Brann}, \citenamefont {Trimmer}, \citenamefont {Sheikh}, \citenamefont
  {Arroyave}, \citenamefont {Reddy}, \citenamefont {Jeong}, \citenamefont
  {Raps} \emph {et~al.}}]{Pfister-2020}%
  \BibitemOpen
  \bibfield  {author} {\bibinfo {author} {\bibfnamefont {P.}~\bibnamefont
  {Pfister}}, \bibinfo {author} {\bibfnamefont {B.~C.}\ \bibnamefont {Smith}},
  \bibinfo {author} {\bibfnamefont {B.~J.}\ \bibnamefont {Evans}}, \bibinfo
  {author} {\bibfnamefont {J.~H.}\ \bibnamefont {Brann}}, \bibinfo {author}
  {\bibfnamefont {C.}~\bibnamefont {Trimmer}}, \bibinfo {author} {\bibfnamefont
  {M.}~\bibnamefont {Sheikh}}, \bibinfo {author} {\bibfnamefont
  {R.}~\bibnamefont {Arroyave}}, \bibinfo {author} {\bibfnamefont
  {G.}~\bibnamefont {Reddy}}, \bibinfo {author} {\bibfnamefont {H.-Y.}\
  \bibnamefont {Jeong}}, \bibinfo {author} {\bibfnamefont {D.~A.}\ \bibnamefont
  {Raps}}, \emph {et~al.},\ }\bibfield  {title} {\bibinfo {title} {Odorant
  receptor inhibition is fundamental to odor encoding},\ }\href@noop {}
  {\bibfield  {journal} {\bibinfo  {journal} {Curr. Biol.}\ }\textbf {\bibinfo
  {volume} {30}},\ \bibinfo {pages} {2574} (\bibinfo {year}
  {2020})}\BibitemShut {NoStop}%
\bibitem [{\citenamefont {Li}\ \emph {et~al.}(2022)\citenamefont {Li},
  \citenamefont {Molday}, \citenamefont {Lin}, \citenamefont {Ren},
  \citenamefont {Fleischmann}, \citenamefont {Molday},\ and\ \citenamefont
  {Yau}}]{li-pnas119-2022}%
  \BibitemOpen
  \bibfield  {author} {\bibinfo {author} {\bibfnamefont {R.-C.}\ \bibnamefont
  {Li}}, \bibinfo {author} {\bibfnamefont {L.~L.}\ \bibnamefont {Molday}},
  \bibinfo {author} {\bibfnamefont {C.-C.}\ \bibnamefont {Lin}}, \bibinfo
  {author} {\bibfnamefont {X.}~\bibnamefont {Ren}}, \bibinfo {author}
  {\bibfnamefont {A.}~\bibnamefont {Fleischmann}}, \bibinfo {author}
  {\bibfnamefont {R.~S.}\ \bibnamefont {Molday}},\ and\ \bibinfo {author}
  {\bibfnamefont {K.-W.}\ \bibnamefont {Yau}},\ }\bibfield  {title} {\bibinfo
  {title} {Low signaling efficiency from receptor to effector in olfactory
  transduction: A quantified ligand-triggered gpcr pathway},\ }\href@noop {}
  {\bibfield  {journal} {\bibinfo  {journal} {Proc. Natl. Aca. Sci., U.S.A.}\
  }\textbf {\bibinfo {volume} {119}},\ \bibinfo {pages} {e2121225119} (\bibinfo
  {year} {2022})}\BibitemShut {NoStop}%
\bibitem [{\citenamefont {Rodieck}(1998)}]{rodieck-98}%
  \BibitemOpen
  \bibfield  {author} {\bibinfo {author} {\bibfnamefont {R.~W.}\ \bibnamefont
  {Rodieck}},\ }\href@noop {} {\emph {\bibinfo {title} {The first steps in
  seeing.}}}\ (\bibinfo  {publisher} {Sinauer Associates},\ \bibinfo {year}
  {1998})\BibitemShut {NoStop}%
\bibitem [{\citenamefont {Reddy}\ \emph
  {et~al.}(2018{\natexlab{b}})\citenamefont {Reddy}, \citenamefont {Zak},
  \citenamefont {Vergassola},\ and\ \citenamefont {Murthy}}]{reddy-2018}%
  \BibitemOpen
  \bibfield  {author} {\bibinfo {author} {\bibfnamefont {G.}~\bibnamefont
  {Reddy}}, \bibinfo {author} {\bibfnamefont {J.~D.}\ \bibnamefont {Zak}},
  \bibinfo {author} {\bibfnamefont {M.}~\bibnamefont {Vergassola}},\ and\
  \bibinfo {author} {\bibfnamefont {V.~N.}\ \bibnamefont {Murthy}},\ }\bibfield
   {title} {\bibinfo {title} {Antagonism in olfactory receptor neurons and its
  implications for the perception of odor mixtures},\ }\href@noop {} {\bibfield
   {journal} {\bibinfo  {journal} {eLife}\ }\textbf {\bibinfo {volume} {7}},\
  \bibinfo {pages} {e34958} (\bibinfo {year} {2018}{\natexlab{b}})}\BibitemShut
  {NoStop}%
\bibitem [{\citenamefont {Xu}\ \emph {et~al.}(2023{\natexlab{b}})\citenamefont
  {Xu}, \citenamefont {Zou},\ and\ \citenamefont {Firestein}}]{xu_2023}%
  \BibitemOpen
  \bibfield  {author} {\bibinfo {author} {\bibfnamefont {L.}~\bibnamefont
  {Xu}}, \bibinfo {author} {\bibfnamefont {D.-J.}\ \bibnamefont {Zou}},\ and\
  \bibinfo {author} {\bibfnamefont {S.}~\bibnamefont {Firestein}},\ }\bibfield
  {title} {\bibinfo {title} {Odor mixtures: A chord with silent notes},\
  }\href@noop {} {\bibfield  {journal} {\bibinfo  {journal} {Front. Ecol.
  Evol.}\ }\textbf {\bibinfo {volume} {11}},\ \bibinfo {pages} {1135486}
  (\bibinfo {year} {2023}{\natexlab{b}})}\BibitemShut {NoStop}%
\bibitem [{\citenamefont {de~March}\ \emph {et~al.}(2020)\citenamefont
  {de~March}, \citenamefont {Titlow}, \citenamefont {Sengoku}, \citenamefont
  {Breheny}, \citenamefont {Matsunami},\ and\ \citenamefont
  {McClintock}}]{deMarch_2020}%
  \BibitemOpen
  \bibfield  {author} {\bibinfo {author} {\bibfnamefont {C.~A.}\ \bibnamefont
  {de~March}}, \bibinfo {author} {\bibfnamefont {W.~B.}\ \bibnamefont
  {Titlow}}, \bibinfo {author} {\bibfnamefont {T.}~\bibnamefont {Sengoku}},
  \bibinfo {author} {\bibfnamefont {P.}~\bibnamefont {Breheny}}, \bibinfo
  {author} {\bibfnamefont {H.}~\bibnamefont {Matsunami}},\ and\ \bibinfo
  {author} {\bibfnamefont {T.~S.}\ \bibnamefont {McClintock}},\ }\bibfield
  {title} {\bibinfo {title} {Modulation of the combinatorial code of odorant
  receptor response patterns in odorant mixtures},\ }\href@noop {} {\bibfield
  {journal} {\bibinfo  {journal} {Mol. Cell. Neurosci.}\ }\textbf {\bibinfo
  {volume} {104}},\ \bibinfo {pages} {103469} (\bibinfo {year}
  {2020})}\BibitemShut {NoStop}%
\bibitem [{\citenamefont {Rospars}\ \emph {et~al.}(2008)\citenamefont
  {Rospars}, \citenamefont {Lansky}, \citenamefont {Chaput},\ and\
  \citenamefont {Duchamp-Viret}}]{rospars-2008}%
  \BibitemOpen
  \bibfield  {author} {\bibinfo {author} {\bibfnamefont {J.-P.}\ \bibnamefont
  {Rospars}}, \bibinfo {author} {\bibfnamefont {P.}~\bibnamefont {Lansky}},
  \bibinfo {author} {\bibfnamefont {M.}~\bibnamefont {Chaput}},\ and\ \bibinfo
  {author} {\bibfnamefont {P.}~\bibnamefont {Duchamp-Viret}},\ }\bibfield
  {title} {\bibinfo {title} {Competitive and noncompetitive odorant
  interactions in the early neural coding of odorant mixtures},\ }\href@noop {}
  {\bibfield  {journal} {\bibinfo  {journal} {J. Neurosci.}\ }\textbf {\bibinfo
  {volume} {28}},\ \bibinfo {pages} {2659} (\bibinfo {year}
  {2008})}\BibitemShut {NoStop}%
\bibitem [{\citenamefont {Bak}\ \emph {et~al.}(2018)\citenamefont {Bak},
  \citenamefont {Jang},\ and\ \citenamefont {Hyeon}}]{bak-2018}%
  \BibitemOpen
  \bibfield  {author} {\bibinfo {author} {\bibfnamefont {J.~H.}\ \bibnamefont
  {Bak}}, \bibinfo {author} {\bibfnamefont {S.~J.}\ \bibnamefont {Jang}},\ and\
  \bibinfo {author} {\bibfnamefont {C.}~\bibnamefont {Hyeon}},\ }\bibfield
  {title} {\bibinfo {title} {Implications for human odor sensing revealed from
  the statistics of odorant-receptor interactions},\ }\href@noop {} {\bibfield
  {journal} {\bibinfo  {journal} {PLOS Comput. Biol.}\ }\textbf {\bibinfo
  {volume} {14}},\ \bibinfo {pages} {e1006175} (\bibinfo {year}
  {2018})}\BibitemShut {NoStop}%
\bibitem [{\citenamefont {Reisert}(2010)}]{Reisert-2010}%
  \BibitemOpen
  \bibfield  {author} {\bibinfo {author} {\bibfnamefont {J.}~\bibnamefont
  {Reisert}},\ }\bibfield  {title} {\bibinfo {title} {Origin of basal activity
  in mammalian olfactory receptor neurons},\ }\href@noop {} {\bibfield
  {journal} {\bibinfo  {journal} {J. Gen. Physiol.}\ }\textbf {\bibinfo
  {volume} {136}},\ \bibinfo {pages} {529} (\bibinfo {year}
  {2010})}\BibitemShut {NoStop}%
\bibitem [{\citenamefont {Mistry}\ \emph {et~al.}(2013)\citenamefont {Mistry},
  \citenamefont {Theodorou}, \citenamefont {Schaal},\ and\ \citenamefont
  {Kawato}}]{Connelly-2013}%
  \BibitemOpen
  \bibfield  {author} {\bibinfo {author} {\bibfnamefont {M.}~\bibnamefont
  {Mistry}}, \bibinfo {author} {\bibfnamefont {E.}~\bibnamefont {Theodorou}},
  \bibinfo {author} {\bibfnamefont {S.}~\bibnamefont {Schaal}},\ and\ \bibinfo
  {author} {\bibfnamefont {M.}~\bibnamefont {Kawato}},\ }\bibfield  {title}
  {\bibinfo {title} {Optimal control of reaching includes kinematic
  constraints},\ }\href@noop {} {\bibfield  {journal} {\bibinfo  {journal} {J.
  Neurophysiol.}\ }\textbf {\bibinfo {volume} {110}},\ \bibinfo {pages} {1}
  (\bibinfo {year} {2013})}\BibitemShut {NoStop}%
\bibitem [{\citenamefont {McLaughlin}\ \emph {et~al.}(1992)\citenamefont
  {McLaughlin}, \citenamefont {McKinnon},\ and\ \citenamefont
  {Margolskee}}]{McLaughlin-1992}%
  \BibitemOpen
  \bibfield  {author} {\bibinfo {author} {\bibfnamefont {S.~K.}\ \bibnamefont
  {McLaughlin}}, \bibinfo {author} {\bibfnamefont {P.~J.}\ \bibnamefont
  {McKinnon}},\ and\ \bibinfo {author} {\bibfnamefont {R.~F.}\ \bibnamefont
  {Margolskee}},\ }\bibfield  {title} {\bibinfo {title} {Gustducin is a
  taste-cell-specific g protein closely related to the transducins},\
  }\href@noop {} {\bibfield  {journal} {\bibinfo  {journal} {Nature}\ }\textbf
  {\bibinfo {volume} {357}},\ \bibinfo {pages} {563} (\bibinfo {year}
  {1992})}\BibitemShut {NoStop}%
\bibitem [{\citenamefont {McCorvy}\ and\ \citenamefont
  {Roth}(2015)}]{McCorvy-2015}%
  \BibitemOpen
  \bibfield  {author} {\bibinfo {author} {\bibfnamefont {J.~D.}\ \bibnamefont
  {McCorvy}}\ and\ \bibinfo {author} {\bibfnamefont {B.~L.}\ \bibnamefont
  {Roth}},\ }\bibfield  {title} {\bibinfo {title} {Structure and function of
  serotonin g protein-coupled receptors},\ }\href@noop {} {\bibfield  {journal}
  {\bibinfo  {journal} {Pharmacol. Ther.}\ }\textbf {\bibinfo {volume} {150}},\
  \bibinfo {pages} {129} (\bibinfo {year} {2015})}\BibitemShut {NoStop}%
\end{thebibliography}
\end{document}